\documentclass[twoside]{article}
\usepackage{qic,epsfig,amsfonts,amsmath,amssymb}
\usepackage[top=30truemm,bottom=30truemm,left=20truemm,right=20truemm]{geometry}

\newtheorem{theo}{Theorem}

\newtheorem{prop}{Proposition}
\renewcommand{\thefootnote}{\fnsymbol{footnote}}  

\begin{document}

\normalsize\textlineskip
\thispagestyle{empty}
\setcounter{page}{1}

\vspace*{0.88truein}
\alphfootnote

\fpage{1}

\centerline{\bf Limit theorems of a two-phase quantum walk with one defect}
\vspace*{0.37truein}
\centerline{\footnotesize
Shimpei Endo\footnote{shimpei.endo@lkb.ens.fr}}
\vspace*{0.015truein}
\centerline{\footnotesize\it Laboratoire Kastler Brossel, Ecole Normale Superieure,}
\baselineskip=10pt
\centerline{\footnotesize\it  24 rue Lhomond, 75231 Paris, France}
\vspace*{10pt}
\centerline{\footnotesize
Takako Endo\footnote{endo.takako@ocha.ac.jp\;(e-mail of the corresponding author)}}
\vspace*{0.015truein}
\centerline{\footnotesize\it Department of Physics, Ochanomizu University}
\baselineskip=10pt
\centerline{\footnotesize\it 2-1-1 Ohtsuka, Bunkyo, Tokyo, 112-0012,
Japan}
\vspace*{10pt}
\centerline{\footnotesize 
Norio Konno\footnote{konno@ynu.ac.jp}}
\vspace*{0.015truein}
\centerline{\footnotesize\it Department of Applied Mathematics, Faculty of Engineering, Yokohama National University}
\baselineskip=10pt
\centerline{\footnotesize\it 79-5 Tokiwadai, Hodogaya, Yokohama, 240-8501, Japan}
\vspace*{10pt}
\centerline{\footnotesize 
Etsuo Segawa\footnote{e-segawa@m.tohoku.ac.jp}}
\vspace*{0.015truein}
\centerline{\footnotesize\it Graduate School of Information Sciences, Tohoku University}
\baselineskip=10pt
\centerline{\footnotesize\it 6-3-09 Aramaki Aza, Aoba, Sendai, Miyagi, 980-8579, Japan}
\vspace*{10pt}
\centerline{\footnotesize 
Masato Takei\footnote{takei@ynu.ac.jp}}
\vspace*{0.015truein}
\centerline{\footnotesize\it Department of Applied Mathematics, Faculty of Engineering, Yokohama National University}
\baselineskip=10pt
\centerline{\footnotesize\it 79-5 Tokiwadai, Hodogaya, Yokohama, 240-8501, Japan}
\vspace*{10pt}

\vspace*{0.225truein}
\publisher{(received date)}{(revised date)}
\vspace*{0.21truein}
\begin{abstract}
We treat a position dependent quantum walk (QW) on the line which 
we assign two different time-evolution operators to positive and negative parts respectively. 
We call the model ``the two-phase QW" here, which has been expected to be a mathematical model of the topological insulator. 
We obtain the stationary and time-averaged limit measures related to localization for the two-phase QW with one defect.
This is the first result on localization for the two-phase QW. 
The analytical methods are mainly based on the splitted generating function of the solution for the eigenvalue problem, 
and the generating function of the weight of the passages of the model. 
In this paper, we call the methods ``the splitted generating function method'' and ``the generating function method'', respectively.
The explicit expression of the stationary measure is asymmetric for the origin, and depends on 
the initial state and the choice of the parameters of the model.
On the other hand, the time-averaged limit measure has a starting point symmetry and localization effect
heavily depends on the initial state and the parameters of the model. Regardless of the strong effect of the initial state and the parameters, the time-averaged limit measure also suggests that localization can be always observed for our two-phase QW.
Furthermore, our results imply that there is an interesting relation between the stationary and time-averaged limit measures when the parameters of the model have specific periodicities, which suggests that there is a possibility that we can analyze localization of the two-phase QW with one defect from the stationary measure.
\end{abstract}

\vspace*{10pt}
\vspace*{3pt}
\communicate{to be filled by the Editorial}

\vspace*{1pt}\textlineskip 
\section{Introduction} 
 Quantum walk (QW) was introduced by several researchers of different fields, such as a quantum probability theory, and quantum information theory \cite{gudder, aharonov}. 
In recent years, as a quantum counterpart of the classical random walk, QW has attracted much attention of various
fields, such as quantum algorithms \cite{kempe,ambainis,shenvi}, probability theory \cite{konno-prob,tate,salmi}, physical systems \cite{shikano,kitagawa}.  
From the viewpoint of asymptotic behavior of QWs, we emphasize that there are two kinds of limit theorems. 
The first one is localization theorem, which we focus on in this paper. Localization is
considered as a typical property of discrete-time QWs \cite{tregenna,inui, segawa}.
Konno \cite{konno_2010} and Konno {\it et al.} \cite{segawa} investigated localization of inhomogeneous discrete-time QWs in one dimension. 
Let $X_{t}$ be the random variable of the position of the walker at time $t$. 
In this paper, we say that the QW starting at the origin shows localization if its time-averaged limit measure is strictly positive, i.e.,

\[\lim_{T\to \infty}\dfrac{1}{T}\sum^{T-1}_{t=0}P(X_{t}=0)>0.\]
\noindent
The other definitions of localization were discussed in \cite{joye,scholz,katsura}. 
The second one is the weak limit theorem for $X_{t}/t$ whose typical expression is described in \cite{segawa}.\\
\indent
For the wide range of interest in two-phase systems in quantum scale \cite{skomski,marton}, QW with two phases can be expected to be attributed to its broad applications to many fields. We call the QW ``the two-phase QW'' in this paper.   
QWs are also naturally the systems of choice to explore fundamental issues in quantum physics.
For instance, Kitagawa {\it et al.} \cite{kitagawa} reported that the experimental realizations of QWs with cold atoms, photons, 
and ions come true non-trivial one-dimensional topological phase including a two phase system.
Hence, we expect that the two-phase QW provides the versatile resources for investigating topological insulators 
which are recently the intense issues of theoretical and experimental approaches as a key to construct quantum computer. 
In spite of such interest in two-phase systems, any limit theorems for the two-phase QW
have not been described mathematically as much as ever.
Now preparing strictly analytical results for the two-phase QW may lead to more profound analyzing for the physical systems.\\
\indent
In $2013$, Konno {\it et al.} \cite{segawa} introduced a method which solves the eigenvalue problem;
\begin{align}
U^{(s)}\Psi=\lambda\Psi\label{2phase-koyutimondai}
\end{align}where
$U^{(s)}$ is an $\infty\times\infty$ unitary matrix, and
$\lambda(\in\mathbb{C})$ and $\Psi$ are the solutions of the eigenvalue problem.
The solutions of the eigenvalue problem lead to a stationary measure which is closely related to localization for some typical QWs in one dimension \cite{segawa}.
We take advantage of the splitted generating function of the solutions, and we call the method ``the splitted generating function method (the SGF method)" in this paper.
Using the SGF method, we derive the stationary measure for the two-phase QW with one defect. 
On the other hand, using the generating function of the weight of the passages, 
we derive the time-averaged limit measure corresponding to localization. Here, we call the method ``the generating function method''.
We see that the two measures agree with each other when the parameters of the model have specific periodicities. 
The Fourier analysis \cite{grimmett} and the stationary phase method \cite{nayak} are useful to study position-independent QWs. 
We can analyze position-dependent QWs using the CGMV method \cite{cantero}, 
however, the CGMV method allows for the general discussion of localization properties, that is, the CGMV method provides the time-averaged limit measure at the origin for the typical one-defect QWs on the line,
however, the analysis is more complicated than the SGF method and the generating function method in general. 
We emphasize that it is the first application of the generating function method to position-dependent QW without symmetry for the origin.

The rest of this article is organized as follows. In Section \ref{main},
we introduce the two-phase QW with one defect, the main target in this paper, and present our main results. 
Then in Section \ref{proof-theo1}, we give the proof of Theorem \ref{stat}. 
Finally, the proof of Theorem \ref{average} is devoted in Section \ref{proof-theo2}. 


\section{Model and the main results}\label{main}
\subsection{Model}
Let the QW be a one-dimensional discrete time model defined by the set of $2\times2$ unitary matrices;
\begin{align}
U_{x}=\begin{bmatrix}
a_{x}&b_{x}\\
c_{x}&d_{x}
\end{bmatrix}\;\;\;(x\in \mathbb{Z}).
\end{align}
The subscript $x$ expresses the position. To consider the time evolution, let us divide $U_{x}$ into two parts as follows;
\[U_{x}=P_{x}+Q_{x}\]
with
\[P_{x}=\begin{bmatrix}
a_{x}&b_{x}\\
0&0
\end{bmatrix},\quad Q_{x}=\begin{bmatrix}
0&0\\
c_{x}&d_{x}
\end{bmatrix}.\]
We should remark that $P_{x}$ and $Q_{x}$ express the left and right steps, respectively.
Here the walker has a state at position $x$ and time $t$, and  the  state can be expressed by a two-dimensional vector
\[\Psi_{t}(x)=\begin{bmatrix}\Psi^{L}_{t}(x)\\ \Psi^{R}_{t}(x) \end{bmatrix}\in\mathbb{C}^{2}.\]
Then the time evolution is determined by the recurrence formula
\[\begin{bmatrix}\Psi^{L}_{t}(x)\\ \Psi^{R}_{t}(x) \end{bmatrix}=\begin{bmatrix}
0&0\\
c_{x}&d_{x}
\end{bmatrix}\begin{bmatrix}\Psi^{L}_{t-1}(x-1)\\ \Psi^{R}_{t-1}(x-1)\end{bmatrix}+\begin{bmatrix}
a_{x}&b_{x}\\
0&0
\end{bmatrix} \begin{bmatrix}\Psi^{L}_{t-1}(x+1)\\ \Psi^{R}_{t-1}(x+1) \end{bmatrix}.\]
In this manuscript, we focus on a one-defect two-phase QW on the line, whose unitary matrices are defined as follows;
\par\indent
\par\noindent
\begin{align}
U_{x}=\left\{ \begin{array}{ll}
U_{+}=\dfrac{1}{\sqrt{2}}\begin{bmatrix}
1 & e^{i\sigma_{+}} \\
e^{-i\sigma_{+}} & -1\\
\end{bmatrix}&(x\geq 1)\\\
&\\
U_{-}=\dfrac{1}{\sqrt{2}}\begin{bmatrix}
1 & e^{i\sigma_{-}} \\
e^{-i\sigma_{-}} & -1\\
\end{bmatrix}&(x\leq -1)\\
&\\
U_{0}=\begin{bmatrix}
1 & 0 \\
0 & -1 \\
\end{bmatrix}& (x=0)
\end{array} \right.
\label{2-phase.def}\end{align}
where $\sigma_{\pm}\in\mathbb{R}$.
\par\indent
\par\noindent
The quantum walker shifts differently in positive and negative parts respectively, and the determinants are independent of the position, that is, $\det(U_{x})=-1$ for $x\in\mathbb{Z}$. Hereafter, we call the model ``the two-phase QW with one defect" for short.
We should notice that if $\sigma_{+}=\sigma_{-}$, the model becomes a one-defect QW which has been so far analyzed in detail \cite{segawa}. Furthermore, if $\sigma_{+}=\sigma_{-}=0$, the model becomes a special case of the QW which has been studied in \cite{endo_san}. We should remark that the two-phase QW has a defect at the origin, which enables us to analyze the model simply. Thus, one of the future problems is to analyze a QW which has two phases and does not have any defect. We will report on the results in the forthcoming paper.

\subsection{The stationary measure}
\label{result1}
We should recall  
\par\indent
\par\noindent
\[U^{(s)}=SU_{x}=\begin{bmatrix}
\ddots&\vdots&\vdots&\vdots&\vdots&\cdots\\
\cdots&O&P_{-1}&O&O&O\cdots\\
\cdots&Q_{-2}&O&P_{0}&O&O\cdots\\
\cdots&O&Q_{-1}&O&P_{1}&O\cdots\\
\cdots&O&O&Q_{0}&O&P_{2}\cdots\\
\cdots&O&O&O&Q_{1}&O\cdots\\
\cdots&\vdots&\vdots&\vdots&\vdots&\ddots
\end{bmatrix}\;\;\;
with\;\;\;O=\begin{bmatrix}0&0\\0&0\end{bmatrix},\]
\par\indent
\par\noindent
and  the quantum walker shifts left or right in accordance with 
$P_{x}$ or $Q_{x}\;(x\in\mathbb{Z})$ \cite{watanabe}, where $x$ is the position of the walker.
Here $S$ is the standard shift operator defined by
\[S=\sum_{x}(|x\rangle\langle x+1|\otimes|L\rangle\langle L|+(|x\rangle\langle x-1|\otimes|R\rangle\langle R|).\]
In this subsection, we present one of our main results, the stationary measure of the two-phase QW with one defect.
From now on, we let $U^{(s)}$ be an $\infty\times\infty$ unitary matrix of the two-phase QW with one defect.
Now let us consider the eigenvalue problem 
\begin{align}U^{(s)}\Psi=\lambda\Psi,\label{2phase-eigenvalue.prob.}\end{align}
where $\lambda\;(\in\mathbb{C})$ is the eigenvalue of $U^{(s)}$ and $\Psi$ is the eigenvector, defined by\\ \[\Psi= {}^T\![\cdots,\Psi^{L}(-1),\Psi^{R}(-1),\Psi^{L}(0),\Psi^{R}(0),\Psi^{L}(1),\Psi^{R}(1),\cdots]\in\mathbb{C}^{\infty},\]
where $T$ means the transpose operation.
First of all, we give the solutions of the eigenvalue problem (\ref{2phase-eigenvalue.prob.}). 
The proof of Proposition \ref{prob-meas.} is provided in Section \ref{proof-theo1}.
\par\indent
\begin{prop}
\label{prob-meas.}Let $\lambda^{(j)}$ be the eigenvalue of the unitary matrix $U^{(s)}$ and $\Psi^{(j)}(x)$ be the eigenvector at $x$ corresponding to $\lambda^{(j)}$, where $j=1,2,3,4$. Put $c\in\mathbb{C}$.
\begin{enumerate}
\item For $\lambda^{(1)}=\dfrac{\cos\sigma+(\sin\sigma+\sqrt{2})i}{\sqrt{3+2\sqrt{2}\sin\sigma}}$ and $\Psi^{(1)}(0)={}^T\![\alpha^{(1)},\beta^{(1)}]=
\dfrac{c}{\sqrt{2}}{}^T\!\left[1,-ie^{-(\sigma_{+}+\sigma_{-})i/2}\right],$ we have
\[\qquad\Psi^{L}(x)=\left\{ \begin{array}{ll}
\dfrac{c}{\sqrt{2}}\left(\dfrac{i}{\sqrt{3+2\sqrt{2}\sin\sigma}}\right)^{x} & (x=1,2,\cdots), \\
\dfrac{c}{\sqrt{2}} & (x=0), \\
\left(1+\dfrac{i}{\sqrt{2}}e^{-\frac{\sigma_{+}+3\sigma_{-}}{2}i}\right)c\left(-\dfrac{i}{\sqrt{3+2\sqrt{2}\sin\sigma}}\right)^{-x} & (x=-1,-2,\cdots). \\
\end{array} \right.\]
\[\Psi^{R}(x)=
\left\{ \begin{array}{ll}
\left(\dfrac{e^{-i\sigma_{+}}}{\sqrt{2}}-ie^{-\frac{\sigma_{+}+\sigma_{-}}{2}i}\right)c\left(\dfrac{i}{\sqrt{3+2\sqrt{2}\sin\sigma}}\right)^{x} & (x=1,2,\cdots), \\
-ie^{-\frac{\sigma_{+}+\sigma_{-}}{2}i}\dfrac{c}{\sqrt{2}} & (x=0), \\
-ie^{-\frac{\sigma_{+}+\sigma_{-}}{2}i}\dfrac{c}{\sqrt{2}}\left(-\dfrac{i}{\sqrt{3+2\sqrt{2}\sin\sigma}}\right)^{-x} & (x=-1,-2,\cdots). \\
\end{array} \right.\]

\item For $\lambda^{(2)}=-\lambda^{(1)}$ and $\Psi^{(2)}(0)=\Psi^{(1)}(0)={}^T\![\alpha^{(1)},\beta^{(1)}],$ we have
\[\quad\Psi^{L}(x)=\left\{ \begin{array}{ll}
\dfrac{c}{\sqrt{2}}\left(-\dfrac{i}{\sqrt{3+2\sqrt{2}\sin\sigma}}\right)^{x} & (x=1,2,\cdots), \\
\dfrac{c}{\sqrt{2}} & (x=0), \\
\left(1+\dfrac{i}{\sqrt{2}}e^{-\frac{\sigma_{+}+3\sigma_{-}}{2}i}\right)c\left(\dfrac{i}{\sqrt{3+2\sqrt{2}\sin\sigma}}\right)^{-x} & (x=-1,-2,\cdots). \\
\end{array} \right.\]
\[\Psi^{R}(x)=
\left\{ \begin{array}{ll}
\left(\dfrac{e^{-i\sigma_{+}}}{\sqrt{2}}-ie^{-\frac{\sigma_{+}+\sigma_{-}}{2}i}\right)c\left(-\dfrac{i}{\sqrt{3+2\sqrt{2}\sin\sigma}}\right)^{x} & (x=1,2,\cdots), \\
-ie^{-\frac{\sigma_{+}+\sigma_{-}}{2}i}\dfrac{c}{\sqrt{2}} & (x=0), \\
-ie^{-\frac{\sigma_{+}+\sigma_{-}}{2}i}\dfrac{c}{\sqrt{2}}\left(\dfrac{i}{\sqrt{3+2\sqrt{2}\sin\sigma}}\right)^{-x} & (x=-1,-2,\cdots). 
\end{array} \right.\]

\item For $\lambda^{(3)}=-\dfrac{\cos\sigma+(\sin\sigma-\sqrt{2})i}{\sqrt{3-2\sqrt{2}\sin\sigma}}$ and $\Psi^{(3)}(0)={}^T\![\alpha^{(3)},\beta^{(3)}]
=\dfrac{c}{\sqrt{2}}{}^T\!\left[1,ie^{-(\sigma_{+}+\sigma_{-})i/2}\right],$ we have
\[\Psi^{L}(x)=\left\{\begin{array}{ll}
\dfrac{c}{\sqrt{2}}\left(\dfrac{i}{\sqrt{3-2\sqrt{2}\sin\sigma}}\right)^{x} & (x=1,2,\cdots), \\
\dfrac{c}{\sqrt{2}} & (x=0), \\
\left(1-\dfrac{i}{\sqrt{2}}e^{-\frac{\sigma_{+}+3\sigma_{-}}{2}i}\right)c\left(-\dfrac{i}{\sqrt{3-2\sqrt{2}\sin\sigma}}\right)^{-x} & (x=-1,-2,\cdots). 
\end{array} \right.\]
\[\Psi^{R}(x)=
\left\{ \begin{array}{ll}
\left(\dfrac{e^{-i\sigma_{+}}}{\sqrt{2}}+ie^{-\frac{\sigma_{+}+\sigma_{-}}{2}i}\right)c\left(\dfrac{i}{\sqrt{3-2\sqrt{2}\sin\sigma}}\right)^{x} & (x=1,2,\cdots), \\
ie^{-\frac{\sigma_{+}+\sigma_{-}}{2}i}\dfrac{c}{\sqrt{2}} & (x=0), \\
ie^{-\frac{\sigma_{+}+\sigma_{-}}{2}i}\dfrac{c}{\sqrt{2}}\left(-\dfrac{i}{\sqrt{3-2\sqrt{2}\sin\sigma}}\right)^{-x} & (x=-1,-2,\cdots). 
\end{array} \right.\]

\item For $\lambda^{(4)}=-\lambda^{(3)}$ and $\Psi^{(4)}(0)=\Psi^{(3)}(0)={}^T\![\alpha^{(3)},\beta^{(3)}],$ we have
\[\Psi^{L}(x)=\left\{\begin{array}{ll}
\dfrac{c}{\sqrt{2}}\left(-\dfrac{i}{\sqrt{3-2\sqrt{2}\sin\sigma}}\right)^{x} & (x=1,2,\cdots), \\
\dfrac{c}{\sqrt{2}} & (x=0), \\
\left(1-\dfrac{i}{\sqrt{2}}e^{-\frac{\sigma_{+}+3\sigma_{-}}{2}}\right)c\left(\dfrac{i}{\sqrt{3-2\sqrt{2}\sin\sigma}}\right)^{-x} & (x=-1,-2,\cdots). 
\end{array} \right.\]
\[\Psi^{R}(x)=
\left\{ \begin{array}{ll}
\left(\dfrac{e^{-i\sigma_{+}}}{\sqrt{2}}+ie^{-\frac{\sigma_{+}+\sigma_{-}}{2}i}\right)c\left(-\dfrac{i}{\sqrt{3-2\sqrt{2}\sin\sigma}}\right)^{x} & (x=1,2,\cdots), \\
ie^{-\frac{\sigma_{+}+\sigma_{-}}{2}i}\dfrac{c}{\sqrt{2}} & (x=0), \\
ie^{-\frac{\sigma_{+}+\sigma_{-}}{2}i}\dfrac{c}{\sqrt{2}}\left(\dfrac{i}{\sqrt{3-2\sqrt{2}\sin\sigma}}\right)^{-x} & (x=-1,-2,\cdots). 
\end{array} \right.\]
\end{enumerate}
\par\indent
\end{prop}
Here, as for the number of the eigenvalues for one-dimensional QWs, it has been known that there are three cases, that is, their unitary matrices have $0, 2,$ or $4$ eigenvalues \cite{mjcantero}.
Noting that the stationary measure at the position $x$ is defined by $\mu(x)=|\Psi^{L}(x)|^{2}+|\Psi^{R}(x)|^{2}$ \cite{watanabe}, we obtain
\par\indent
\begin{theo}
\label{stat}
\begin{enumerate}
\item For $\lambda^{(1)}$ and $\Psi^{(1)}(0)$, and $\lambda^{(2)}$ and $\Psi^{(2)}(0)$, we have
\[\mu(x)=\left\{ \begin{array}{ll}
(2+\sqrt{2}\sin\sigma)|c|^{2}\left(\dfrac{1}{3+2\sqrt{2}\sin\sigma}\right)^{x} & (x=1,2,\cdots), \\
|c|^{2} & (x=0),\\
\left\{2+\sqrt{2}\sin\left(\dfrac{\sigma_{+}+3\sigma_{-}}{2}\right)\right\}|c|^{2}\left(\dfrac{1}{3+2\sqrt{2}\sin\sigma}\right)^{-x}&(x=-1,-2,\cdots).
\end{array} \right.\]

\item  For $\lambda^{(3)}$ and  $\Psi^{(3)}(0)$, and $\lambda^{(4)}$ and  $\Psi^{(4)}(0)$, we have

\[\mu(x)=\left\{ \begin{array}{ll}
(2-\sqrt{2}\sin\sigma)|c|^{2}\left(\dfrac{1}{3-2\sqrt{2}\sin\sigma}\right)^{x} & (x=1,2,\cdots), \\
|c|^{2} & (x=0),\\
\left\{2-\sqrt{2}\sin\left(\dfrac{\sigma_{+}+3\sigma_{-}}{2}\right)\right\}|c|^{2}\left(\dfrac{1}{3-2\sqrt{2}\sin\sigma}\right)^{-x}&(x=-1,-2,\cdots).
\end{array} \right.\]
\end{enumerate}
\end{theo}

\par\indent
\par\noindent
We should note that $\sum_{x\in\mathbb{Z}}\mu(x)$ strongly depends on $c(\in\mathbb{C})$, and choosing appropriate $c$, then, \[\sum_{x\in\mathbb{Z}}\mu(x)=1\] holds.   
Our explicit expression of the stationary measure has an exponential decay, where the decay rate depends on $\lambda^{(j)}\;(j=1,2,3,4)$, 
however, the stationary measure has the same decay rate in both $x\geq1$ and $x\leq-1$ for each case. On the other hand, the coefficients depend on both the position and $\lambda^{(j)}\;(j=1,2,3,4)$.
 Moreover, the results indicate that the stationary measure strongly depends on the parameters of the model, and is not symmetric for the origin.

\subsection{The time-averaged limit measure}
\label{time}
Localization of the QW with one defect is defined by the time-averaged limit measure \cite{endo}. 
In this subsection, we give the time-averaged limit measure for the two-phase QW with one defect, another main result of our study, by using the generating functions of the weight of the passages.
Now, we have the following limit theorem with respect to localization for the two-phase QW with one defect starting at the origin with the initial state $\varphi_{0}={}^T\![\alpha,\beta]$, where $\alpha,\beta\in\mathbb{C}$. Let  $\overline{\mu}_{\infty}(x)$ be the time-averaged limit measure of the two-phase QW with one defect.
\par\indent
\begin{theo}
\label{average}
\noindent 
Put $\sigma=(\sigma_{+}-\sigma_{-})/2$, $\tilde{\sigma}=(\sigma_{+}+\sigma_{-})/2$ and the polar displays $\alpha={}^T\!(a,\phi_{1}),\;\beta={}^T\!(b,\phi_{2})$ with $a,b\geq0$ and $a^{2}+b^{2}=1$. Note $\tilde{\phi}_{12}=\phi_{1}-\phi_{2}\;$.
Then, we have 
\begin{align}\overline{\mu}_{\infty}(x)=I_{\{-1/\sqrt{2}\leq\sin\sigma\leq 1\}}(x)\nu^{(+)}(x;\sigma)+I_{\{-1\leq\sin\sigma\leq1/\sqrt{2}\}}(x)\nu^{(-)}(x;\sigma),\label{2-phase.timeaveraged}\end{align}

where
\[\nu^{(\pm)}(x;\sigma)=\left(\dfrac{1\pm\sqrt{2}\sin\sigma}{3\pm2\sqrt{2}\sin\sigma}\right)^{2}\{1\mp2\Re{(ie^{-i\tilde{\sigma}}\alpha\overline{\beta})}\}\left\{\delta_{0}(x)+(1-\delta_{0}(x))(2\pm\sqrt{2}\sin\sigma)\left(\dfrac{1}{3\pm2\sqrt{2}\sin\sigma}\right)^{|x|}\right\}.\]
Here,
\[I_{A}(x)=\left\{ \begin{array}{ll}
1 & (x\in A) \\
0 & (x\notin A) \\
\end{array} \right..\]
\end{theo}
\par\indent
\par\noindent
We emphasize that the time-averaged limit measure has an origin symmetry and an exponential decay for the position, and localization heavily depends on parameter $\sigma$ and the initial state. The analytical expression also implies that there is a possibility of localizing if $\sigma_{+}\neq\sigma_{-}$. \\
\indent
Furthermore, comparing Theorem \ref{stat} with Theorem \ref{average}, there seems to be corresponding relationships between the eigenvalues $\lambda^{(j)}\;(j=1,2,3,4)$ in Theorem \ref{stat} and the range of $\sin\sigma$ as follows;
\begin{align}\lambda^{(1)},\;\lambda^{(2)}\longleftrightarrow\;-1/\sqrt{2}\leq\sin\sigma\leq 1: \nu^{(+)}(x;\sigma),\label{assumption1}\end{align}
and
\begin{align}\lambda^{(3)},\;\lambda^{(4)}\longleftrightarrow\;-1\leq\sin\sigma\leq1/\sqrt{2}: \nu^{(-)}(x;\sigma),\label{assumption2}\end{align}
Here, if one of the conditions\\

\hspace{20mm}$\left\{
\begin{array}{l}
\overline{\mu}_{\infty}(0)=|c|^{2},\\
\\
\sigma_{-}=n\pi\quad(n\in\mathbb{Z}),\\
\end{array}
\right.$\hspace{15mm}or\hspace{15mm}$\left\{
\begin{array}{l}
\overline{\mu}_{\infty}(0)=|c|^{2},\\
\\
\sigma_{+}+\sigma_{-}=(2n+1)\pi\quad(n\in\mathbb{Z}),\\
\end{array}
\right.$\\
\par
\noindent
is satisfied and assuming \eqref{assumption1} and  \eqref{assumption2}, then, the parts  $\nu^{(\pm)}(x;\sigma)$ of the time-averaged limit measure and stationary measure agree with each other depending on the eigenvalues $\lambda^{(j)}\;(j=1,2,3,4)$ in Theorem \ref{stat}.

In addition, the two main results suggest that the SGF method gives more tight conditions for localization than the generating function method.
We should remark that we can not see the probability distribution only from 
the stationary or the time-averaged limit measures, since $\sum_{x\in\mathbb{Z}}\overline{\mu}_{\infty}(x)<1$ holds. 
Here to clarify the relation between the two-phase QW and the topological insulator in concrete is one of the interesting future problems.
Section $4$ is devoted to the proof of  Theorem $2$.


\subsection{Examples}
\label{examples}
Here in order to see the relation between the stationary and time-averaged limit measures and to grasp what our analytical results suggest, we consider two simple examples.
\begin{enumerate}
\item At first, we see the QW whose time evolution is determined by
\[U_{x}=\left\{ \begin{array}{ll}
\dfrac{1}{\sqrt{2}}\begin{bmatrix}1 & 1\\ 1 & -1 \end{bmatrix}& (x\neq 0), \\
&\\
\begin{bmatrix}1 & 0\\ 0 & -1 \end{bmatrix} & (x=0),
\end{array} \right.\]
which is obtained by putting $\sigma_{+}=\sigma_{-}=0$ in Eq. \eqref{2-phase.def}.\\

Now Theorem \ref{stat} gives the stationary measure as follows;\\
For $\lambda^{(1)}=\dfrac{1}{\sqrt{3}}(1+\sqrt{2}i),\;\lambda^{(2)}=-\lambda^{(1)}$ and $\Psi^{(1)}(0)=\Psi^{(2)}(0)=\dfrac{c}{\sqrt{2}}{}^T\![1,- i]$, and $\lambda^{(3)}=-\dfrac{1}{\sqrt{3}}(1-\sqrt{2}i),\;\lambda^{(4)}=-\lambda^{(3)}$ and $\Psi^{(3)}(0)=\Psi^{(4)}(0)=\dfrac{c}{\sqrt{2}}{}^T\![1, i]$, the stationary measure can be expressed by
\begin{align}
\mu(x)=\left\{ \begin{array}{ll}
2|c|^{2}\left(\dfrac{1}{3}\right)^{|x|} & (x=\pm1,\pm2,\cdots),\\
|c|^{2}& (x=0).
\end{array} \right.\label{exampleteijyo1-0}
\end{align}
On the other hand, Theorem \ref{average} gives the time-averaged limit measure by
\begin{align*}\overline{\mu}_{\infty}(x)&=\nu^{(+)}(x;0)+\nu^{(-)}(x;0)\\
&=\dfrac{1}{9}\{1-2\Re{(i\alpha\overline{\beta})}\}\times\left\{\delta_{0}(x)+2(1-\delta_{0}(x))\left(\dfrac{1}{3}\right)^{|x|}\right\}+\\
&\dfrac{1}{9}\{1+2\Re{(i\alpha\overline{\beta})}\}\times\left\{\delta_{0}(x)+2(1-\delta_{0}(x))\left(\dfrac{1}{3}\right)^{|x|}\right\},\end{align*}
which agrees with the expression obtained by Theorem $3.2$ in Ref. \cite{segawa}. Here,
\[\nu^{(\pm)}(x;0)=\dfrac{1}{9}\{1\mp2\Re{(i\alpha\overline{\beta})}\}\times\left\{\delta_{0}(x)+2(1-\delta_{0}(x))\left(\dfrac{1}{3}\right)^{|x|}\right\}.\]
Thereby, putting $|c|=\sqrt{1-2\Re{(i\alpha\overline{\beta})}}/3$ in Eq. \eqref{exampleteijyo1-0}, and taking into account of the assumption \eqref{assumption1}, the stationary measure for $\lambda^{(1)}$ (or $\lambda^{(2)}$) and $\nu^{(+)}(x;0)$ coincide with each other.  Letting $|c|=\sqrt{1+2\Re{(i\alpha\overline{\beta})}}/3$ in Eq. \eqref{exampleteijyo1-0}, the stationary measure for $\lambda^{(3)}$ (or $\lambda^{(4)}$) and $\nu^{(-)}(x;0)$ also agree with each other.
We also see that localization happens for the model, for we have $\overline{\mu}_{\infty}(0)=2/9 (>0)$.\\
\item
Next we consider the QW defined by
\begin{align}
U_{x}=\left\{ \begin{array}{ll}
\dfrac{1}{\sqrt{2}}\begin{bmatrix}1 & -i\\ i & -1 \end{bmatrix}& (x=1,2,\cdots), \\
&\\
\dfrac{1}{\sqrt{2}}\begin{bmatrix}1 & -1\\ -1 & -1 \end{bmatrix} & (x=-1,-2,\cdots),\\
&\\
\begin{bmatrix}1 & 0\\ 0 & -1 \end{bmatrix} & (x=0).
\end{array} \right.\label{examplemode}
\end{align}  
We obtain the QW by putting $\sigma_{+}=3\pi/2$ and $\sigma_{-}=\pi$ in Eq. \eqref{2-phase.def}.\\

\indent
Using Theorems \ref{stat} and \ref{average}, we obtain the stationary and time-averaged limit measures in the following;
For $\lambda^{(1)}=\dfrac{1}{\sqrt{10}}(1+3i),\;\lambda^{(2)}=-\lambda^{(1)}$ and $\Psi^{(1)}(0)=\Psi^{(2)}(0)=\dfrac{c}{\sqrt{2}}{}^T\![1, \dfrac{1}{\sqrt{2}}(1+i)]$, Theorem \ref{stat} yields the stationary measure by
\begin{align}
\mu(x)=\left\{ \begin{array}{ll}
3|c|^{2}\left(\dfrac{1}{5}\right)^{|x|} & (x=\pm1,\pm2,\cdots),\\
|c|^{2}& (x=0).
\end{array} \right.\label{exampleteijyo1-1}
\end{align}
On the other hand, Theorem \ref{average} gives
\[ \overline{\mu}_{\infty}(x)=\nu^{(+)}(x;\pi/4)=\dfrac{4}{25}\{1-2\Re{(ie^{-i\frac{5\pi}{4}}\alpha\overline{\beta})}\}\times\left\{\delta_{0}(x)+3(1-\delta_{0}(x))\left(\dfrac{1}{5}\right)^{|x|}\right\}.\]
Therefore, if we put $|c|=2\sqrt{1-2\Re{(ie^{-i5\pi/4}\alpha\overline{\beta})}}/5$ in Eq. \eqref{exampleteijyo1-1}, and combining with the assumption \eqref{assumption1}, $\nu^{(+)}(x;\pi/4)$ and the stationary measure for $\lambda^{(1)}$ (or $\lambda^{(2)}$) agree with each other. 
On the other hand, for 
$\lambda^{(3)}=-\dfrac{1}{\sqrt{2}}+\dfrac{i}{\sqrt{2}},\;\lambda^{(4)}=-\lambda^{(3)}$ and $\Psi^{(3)}(0)=\Psi^{(4)}(0)=\dfrac{c}{\sqrt{2}}{}^T\![1, -\dfrac{1}{\sqrt{2}}(1+i)]$, Theorem \ref{stat} yields
\begin{align}
\mu(x)=|c|^{2}.\label{exampleteijyo1-2}
\end{align}
On the other hand, Theorem \ref{average} gives
\[\nu^{(-)}(x;\pi/4)=0.\]
Therefore, if we put $|c|=0$ in Eq. \eqref{exampleteijyo1-2}, then, $\nu^{(-)}(x;\pi/4)$ and the stationary measure for $\lambda^{(3)}$ (or $\lambda^{(4)}$) coincide with each other.\\
We also see that $ \overline{\mu}_{\infty}(0)=\dfrac{4}{25}\{1-2\Re{(ie^{-i5\pi/4}\alpha\overline{\beta})}\}$, and localization can be occured by appropriate choice of the initial state.\\
\par\indent
Here, we show two kinds of the numerical results. One is the time-average of the probability distribution for parameter $(\sigma_{+},\sigma_{-})=(3\pi/2, \pi)$ for two initial states ${}^T\![\alpha,\beta]= {}^T\![1,0]$ and ${}^T\![i/\sqrt{2}, 1/\sqrt{2}]$ at time $t=100$, $1000$, and $10000$ (Figs. \ref{fig1},\ref{fig2}). The other is the probability distribution at time $10000$ (Figs. \ref{fig3},\ref{fig4}).
From Figs. \ref{fig1} and \ref{fig2}, we see that each numerical  result coincides with the analytical result for $|x|\ll\log(t)$, which indicates that the numerical results gradually close to the analytical result at very low speed.
We emphasize that the analytical result of the time-averaged limit measure has an origin symmetry, however, Figs. \ref{fig3} and \ref{fig4} indicate that the probability distributions do not have the symmetry. We remark that the asymmetry of the probability distributions can be expressed by the weak limit theorem for $X_{t}/t$, which we will report in the forthcoming paper.

\begin{figure}[h]
\begin{minipage}{0.5\hsize}
\centerline{\epsfig{file=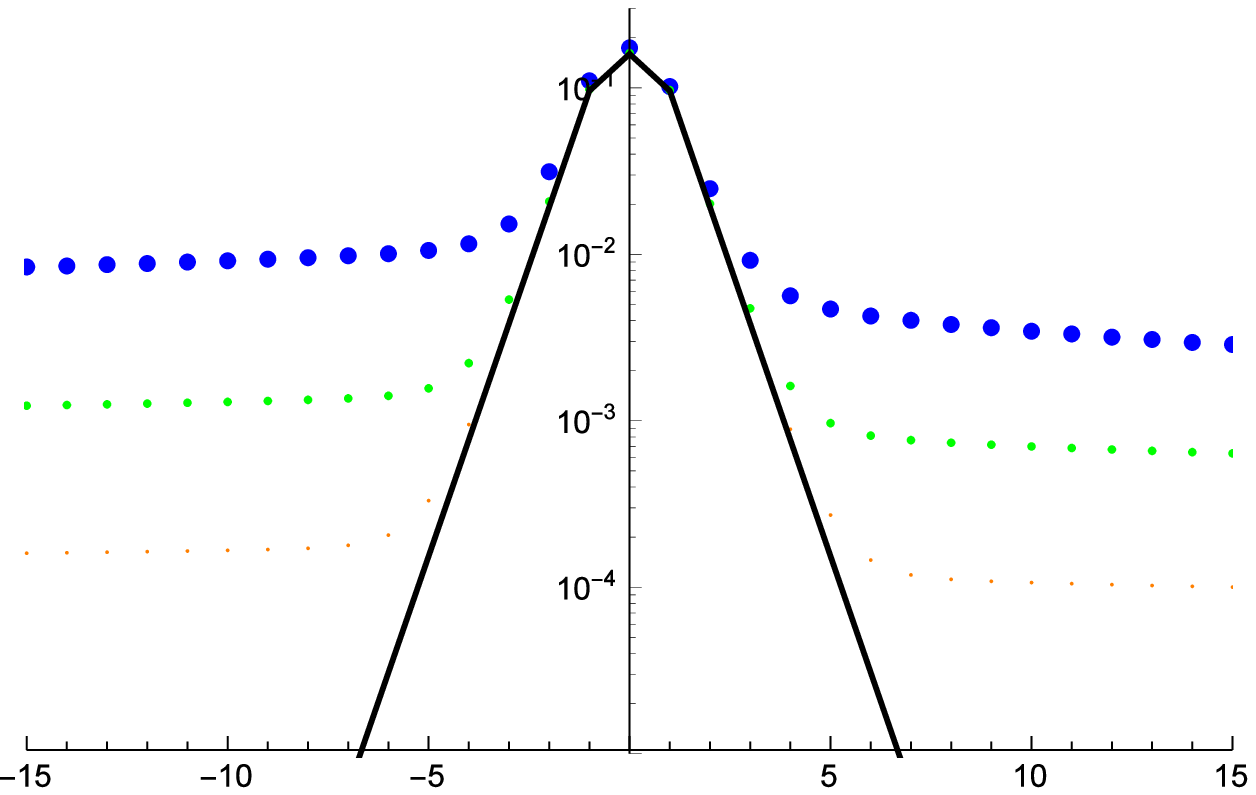, width=7.1cm}} 
\vspace*{13pt}
\fcaption{\label{fig1} ${}^T\![\alpha,\beta]= {}^T\![1,0]$ case.\\
Time-average of the probability at time $t=100$ (blue points),\\ 
 $t=1000$ (green points), and  $t=10000$ (orange points). \\
 Black points and curve denotes the time-averaged limit measure \\in
 Eq. \eqref{2-phase.timeaveraged} }
\end{minipage}\quad\quad
\begin{minipage}{0.5\hsize}
\centerline{\epsfig{file=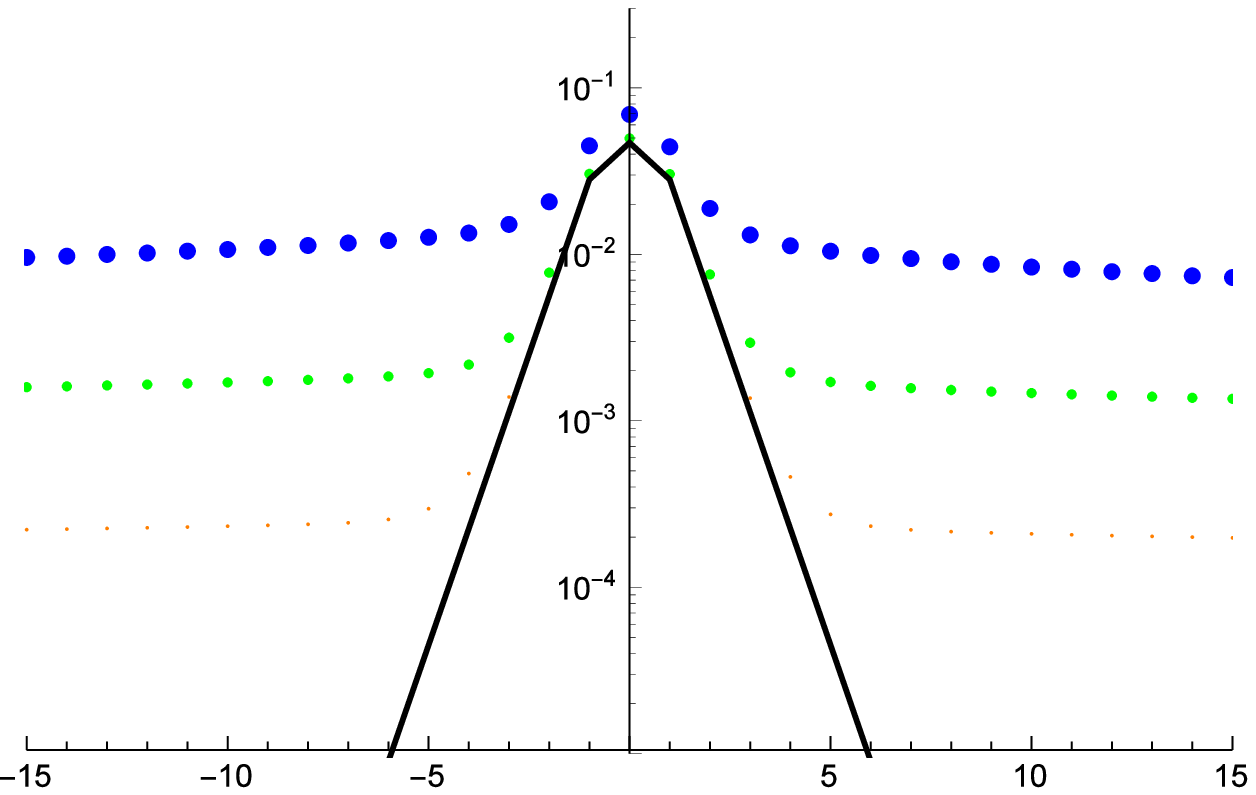, width=7.1cm}} 
\vspace*{13pt}
\fcaption{\label{fig2}${}^T\![\alpha,\beta]={}^T\![i/\sqrt{2}, 1/\sqrt{2}]$ case.\\
Time-average of the probability at time $t=100$ (blue points), \\
 $t=1000$ (green points), and  $t=10000$ (orange points). \\
 Black points and curve denotes the time-averaged limit measure \\in
 Eq. \eqref{2-phase.timeaveraged} 
}
 \end{minipage}
\end{figure}

\begin{figure}[h]
\begin{minipage}{0.5\hsize}
\centerline{\epsfig{file=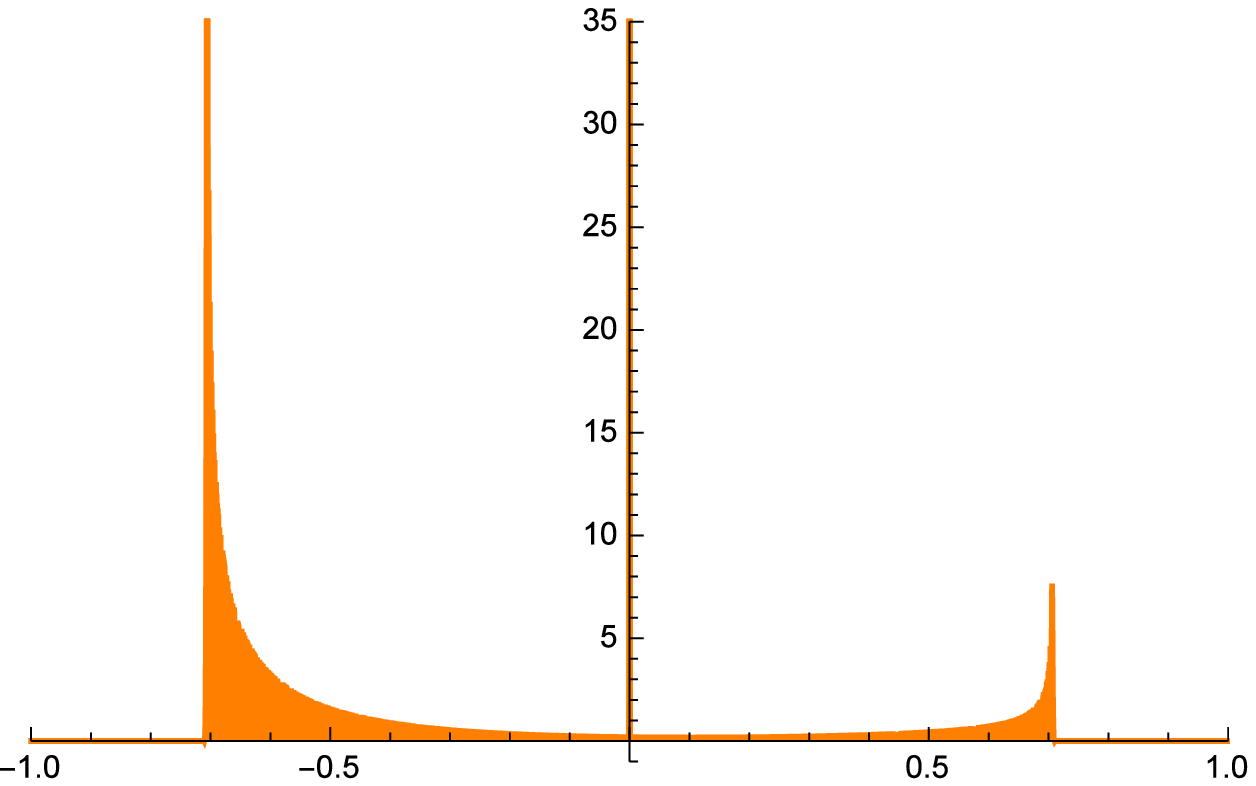, width=7.1cm}} 
\vspace*{13pt}
\fcaption{\label{fig3} ${}^T\![\alpha,\beta]= {}^T\![1,0]$ case.\\
Probability distribution at time $10000$.} 
\end{minipage}
\begin{minipage}{0.5\hsize}
\centerline{\epsfig{file=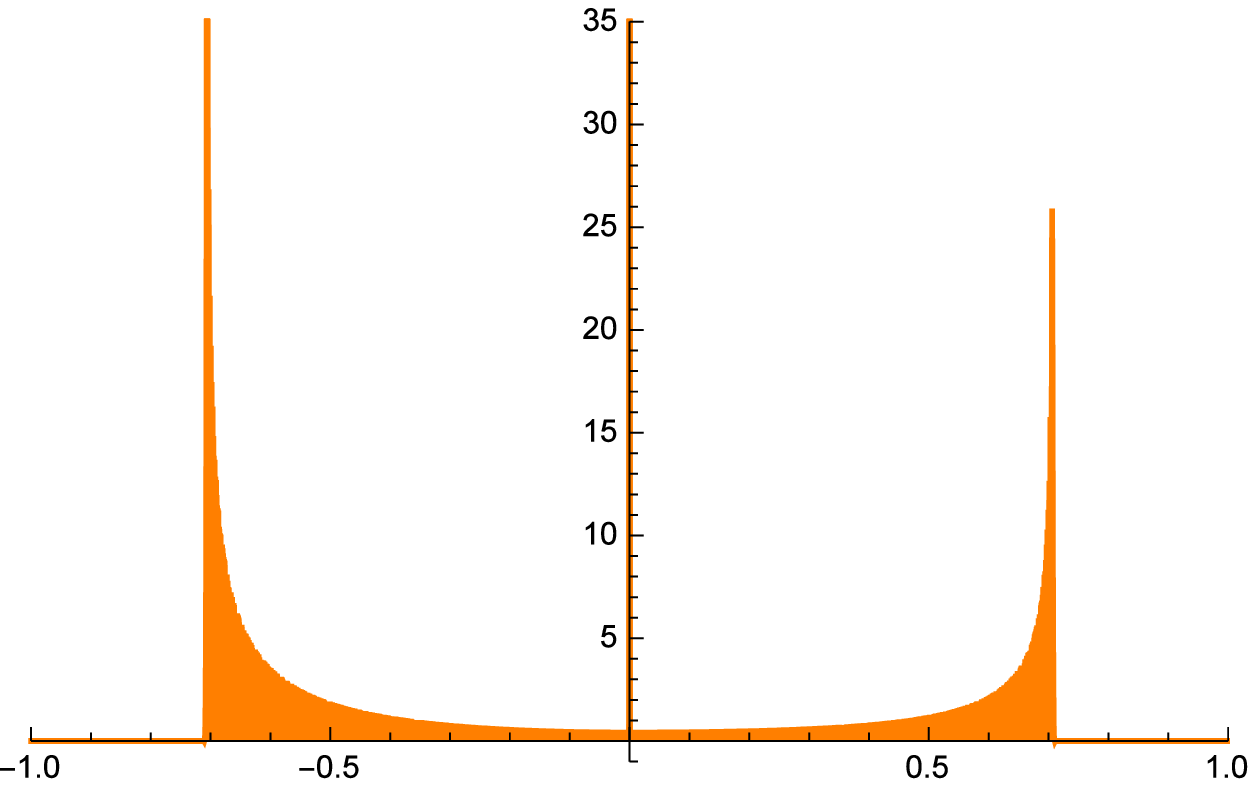, width=7.1cm}} 
\vspace*{13pt}
\fcaption{\label{fig4}${}^T\![\alpha,\beta]={}^T\![i/\sqrt{2}, 1/\sqrt{2}]$ case.\\
Probability distribution at time $10000$.
}
 \end{minipage}
\end{figure}

\end{enumerate}


\section{Proof of Theorem \ref{stat}}
\label{proof-theo1}
Let us begin with the eigenvalue problem
\begin{align}U^{(s)}\Psi=\lambda\Psi,\label{eigenvalueproblem}\end{align}
where $\lambda\in\mathbb{C}$ with $|\lambda|=1$.
From now on, we solve the eigenvalue problem (\ref{eigenvalueproblem}) taking advantage of the SGF method.
Rewriting the eigenvalue problem for position $x\in\mathbb{Z}$, we get
\begin{align}
\lambda\Psi(x)=P_{x+1}\Psi(x+1)+Q_{x-1}\Psi(x-1).
\label{bisyou}
\end{align}
Eq. \eqref{bisyou} can be expressed depending on each position as follows;
\begin{enumerate}
\item Case of $x=\pm2,\pm3,\cdots$:\\
\begin{eqnarray}
\lambda\Psi^{L}(x)\!\!\!&=&\!\!\!\dfrac{1}{\sqrt{2}}\Psi^{L}(x+1)+\dfrac{e^{i\sigma_{\pm}}}{\sqrt{2}}\Psi^{R}(x+1),\label{2phase-koyuuti-relation2.1}\\
\lambda\Psi^{R}(x)\!\!\!&=&\!\!\!\dfrac{e^{-i\sigma_{\pm}}}{\sqrt{2}}\Psi^{L}(x-1)-\dfrac{1}{\sqrt{2}}\Psi^{R}(x-1).
\label{2phase-koyuuti-relation2.2}\end{eqnarray}
\item Case of $x=1$:\\
\begin{eqnarray}
\lambda\Psi^{L}(1)\!\!\!&=&\!\!\!\dfrac{1}{\sqrt{2}}\Psi^{L}(2)+\dfrac{e^{i\sigma_{+}}}{\sqrt{2}}\Psi^{R}(2),\label{2phase-koyuuti-relation3.1}\\
\lambda\Psi^{R}(1)\!\!\!&=&\!\!\!-\Psi^{R}(0).
\label{2phase-koyuuti-relation3.2}\end{eqnarray}
\item Case of $x=-1$:\\
\begin{eqnarray}
\lambda\Psi^{L}(-1)\!\!\!&=&\!\!\!\Psi^{L}(0),\label{2phase-koyuuti-relation4.1}\\
\lambda\Psi^{R}(-1)\!\!\!&=&\!\!\!\dfrac{e^{-i\sigma_{-}}}{\sqrt{2}}\Psi^{L}(-2)-\dfrac{1}{\sqrt{2}}\Psi^{R}(-2).
\label{2phase-koyuuti-relation4.2}\end{eqnarray}
\item Case of $x=0$:
\begin{eqnarray}
\lambda\Psi^{L}(0)\!\!\!&=&\!\!\!\dfrac{1}{\sqrt{2}}\Psi^{L}(1)+\dfrac{e^{i\sigma_{+}}}{\sqrt{2}}\Psi^{R}(1),\label{2phase-koyuuti-relation5.1}\\
\lambda\Psi^{R}(0)\!\!\!&=&\!\!\!\dfrac{e^{-i\sigma_{-}}}{\sqrt{2}}\Psi^{L}(-1)-\dfrac{1}{\sqrt{2}}\Psi^{R}(-1).
\label{2phase-koyuuti-relation5.2}\end{eqnarray}

\end{enumerate}
Here, we introduce the generating functions of $\Psi^{j}(x)\;(j=L,R)$ by
\begin{eqnarray}
f^{j}_{+}(z)=\sum^{\infty}_{x=1}\Psi^{j}(x)z^{x},\quad
f^{j}_{-}(z)=\sum^{-\infty}_{x=-1}\Psi^{j}(x)z^{x}.
\label{2phase-bokansuu}\end{eqnarray}
Then, we obtain
\par\indent
\par\noindent
\begin{lemma}
\label{2phase-hodai1}
Put
\begin{eqnarray*}A_{\pm}\!\!\!&=&\!\!\!\begin{bmatrix}\lambda-\dfrac{1}{\sqrt{2}z}&-\dfrac{e^{i\sigma_{\pm}}}{\sqrt{2}z}\nonumber\\\
\dfrac{e^{-i\sigma_{\pm}}}{\sqrt{2}}z&-\lambda-\dfrac{z}{\sqrt{2}}
\end{bmatrix},
\;{\bf f}_{\pm}(z)=\left[\begin{array}{c}f^{L}_{\pm}(z)\\f^{R}_{\pm}(z)\end{array}\right],\nonumber\\\
{\bf a}_{+}(z)\!\!\!&=&\!\!\!\left[\begin{array}{c}-\lambda\alpha\\ \beta z\end{array}\right],\;
{\bf a}_{-}(z)=\left[\begin{array}{c}z^{-1}\alpha\\\lambda\beta\end{array}\right].\end{eqnarray*}
\par\indent
\par\noindent
Then, 
\begin{align}
A_{\pm}{\bf f}_{\pm}(z)={\bf a}_{\pm}(z)\label{2phase-hodai1-houteisiki}
\end{align}
holds.
\end{lemma}

\par\indent
\par\noindent
Noting
\begin{align}
\det A_{\pm}  =-\dfrac{\lambda }{\sqrt{2}z}\left\{z^{2}-\sqrt{2}\left(\dfrac{1}{\lambda}-\lambda\right)z-1\right\},
\label{deta12sou}
\end{align}
we take $\theta_{s}, \> \theta_{l} \in\mathbb{C}$ satisfying
\begin{align}
\det A_{\pm} =-\dfrac{\lambda}{ \sqrt{2}z}(z-\theta_{s})(z-\theta_{l})
\label{deta22sou}
\end{align}
and $|\theta_{s}|\leq1\leq|\theta_{l}|$.
Here, Eqs. \eqref{deta12sou} and \eqref{deta22sou} give $\theta_{s}\theta_{l}=-1$.\\
\\
From now on, let us derive $f_{\pm}^{L}(z)$ and $f_{\pm}^{R}(z)$ from Lemma \ref{2phase-hodai1}.

\begin{enumerate}
\item $f_{+}^{L}(z)$ case. Eq. \eqref{2phase-hodai1-houteisiki} gives
\begin{align*}
f^{L}_{+}(z)
= \dfrac{\lambda \alpha}{\sqrt{2} \det A_+} \left(z + \dfrac{\sqrt{2}\lambda^{2}\alpha +  e^{i \sigma_{+}} \beta}{\lambda\alpha} \right).
\end{align*}
Putting $\theta_{s}=- \dfrac{\sqrt{2}\lambda^{2}\alpha +  e^{i \sigma_{+}} \beta}{\lambda\alpha}$, we have
\begin{align*}
f^{L}_{+}(z)
&= -\dfrac{\alpha z}{z-\theta_{l}} \\
&=
\dfrac{\alpha\theta_{l}^{-1}z}{1-\theta_{l}^{-1}z}
\\
&= -\alpha(\theta_{s}z) 
\left\{ 1+(-\theta_{s}z)+(-\theta_{s}z)^{2}+(-\theta_{s}z)^{3}+\cdots \right\}.
\end{align*}
Hence, we see 
\begin{align}
f^{+}_{L}(z)=\alpha\sum_{x=1}^{\infty}(-\theta_{s}z)^{x}.
\label{panda12sou}
\end{align}
Equation \eqref{panda12sou} and the definition of $f_{+}^{L}(z)$ give
\begin{align}
\Psi^{L}(x)=\alpha(-\theta_{s})^{x}\;\;\;(x=1,2,\cdots),
\end{align}
where
\begin{align}
\theta_{s}=- \dfrac{\sqrt{2}\lambda^{2}\alpha +  e^{i \sigma_{+}} \beta}{\lambda\alpha}.
\label{onene12sou}
\end{align}
\item $f_{+}^{R}(z)$ case. 
Putting $\theta_{s}= \dfrac{ \beta}{ \lambda(e^{-i \sigma_{+}} \alpha + \sqrt{2} \beta)}$, we have from Eq. \eqref{2phase-hodai1-houteisiki}
\begin{align}
f^{R}_{+}(z)
&=- \dfrac{(e^{-i \sigma_{+}} \alpha + \sqrt{2} \beta)\theta_{l}^{-1} z}{1-\theta_{l}^{-1}z}
\nonumber
\\
&=(e^{-i \sigma_{+}} \alpha + \sqrt{2} \beta)\sum_{x=1}^{\infty}(-\theta_{s}z)^{x}.
\label{panda22sou}
\end{align}
Equation \eqref{panda22sou} and the definition of $f_{+}^{R}(z)$ give
\begin{align}
\Psi^{R}(x)=(e^{-i \sigma_{+}} \alpha + \sqrt{2} \beta)(-\theta_{s})^{x}\;\;\;(x=1,2,\cdots),
\end{align}
where
\begin{align}
\theta_{s} =\dfrac{ \beta}{ \lambda(e^{-i \sigma_{+}} \alpha + \sqrt{2} \beta)}.
\label{onene22sou}
\end{align}
\item $f_{-}^{L}(z)$ case.
Putting $\theta_{l}^{-1} = - \dfrac{\alpha }{\lambda(\beta e^{i \sigma_{-}}-\sqrt{2}\alpha)}$,
Eq. \eqref{2phase-hodai1-houteisiki} gives
\begin{align*}
f^{L}_{-}(z)
&=\dfrac{(\sqrt{2}\alpha-\beta e^{i \sigma_{-}})\theta_{s}z^{-1}}{1-\theta_{s}z^{-1}}
\\
&=(\sqrt{2}\alpha-\beta e^{i \sigma_{-}})\left\{\dfrac{\theta_{s}}{z}+\left(\dfrac{\theta_{s}}{z}\right)^{2}+\cdots\right\}.
\end{align*}
Hence, we have
\begin{align}
f^{L}_{-}(z)=(\sqrt{2}\alpha-\beta e^{i \sigma_{-}})\sum^{-\infty}_{x=-1}(\theta^{-1}_{s}z)^{x}.
\label{panda32sou}
\end{align}
Equation \eqref{panda32sou} and the definition of $f^{L}_{-}(z)$ yield
\begin{align*}
\Psi^{L}(x)=(\sqrt{2}\alpha-\beta e^{i \sigma_{-}})\theta_{s}^{-x}\;\;\;(x=-1,-2,\cdots),
\end{align*}
where
\begin{align}
\theta_{s}= \dfrac{\alpha}{\lambda(\sqrt{2}\alpha-\beta e^{i \sigma_{-}})}.
\label{onene32sou}
\end{align}
\item $f_{-}^{R}(z)$ case. Putting
$
\theta_{l}^{-1}= \dfrac{\sqrt{2} \lambda^{2} \beta-e^{-i\sigma_{-}}\alpha}{\lambda\beta},
$
Eq. \eqref{2phase-hodai1-houteisiki} yields

\begin{align}
f^{R}_{-}(z)=\beta\sum^{-\infty}_{x=-1}(\theta_{s}^{-1}z)^{x}.
\label{panda42sou}
\end{align}
Therefore, Eq. \eqref{panda42sou} and the definition of  $f^{R}_{-}(z)$ give
\begin{align*}
\Psi^{R}(x)=\beta\theta_{s}^{-x}\;\;\;(x=-1,-2,\cdots),
\end{align*}
where
\begin{align}
\theta_{s}=\dfrac{e^{-i\sigma_{-}}\alpha-\sqrt{2} \lambda^{2} \beta}{\lambda\beta}.
\label{onene42sou}
\end{align}
\end{enumerate} 
As a result, we obtain
\begin{align}
\Psi(x)=\left\{ \begin{array}{ll}
(-\theta_{s})^{x}
\begin{bmatrix}
\alpha \\ e^{-i \sigma_{+}} \alpha +\sqrt{2} \beta
\end{bmatrix} &(x=1,2,\ldots),\\
\begin{bmatrix}
\alpha \\ 
\beta \end{bmatrix} &(x=0),\\
(\theta_{s})^{|x|}
\begin{bmatrix}
\sqrt{2}\alpha-e^{i \sigma_{-}} \beta\\ \beta 
\end{bmatrix} &(x=-1,-2,\ldots).
\end{array} \right.
\label{araisan2sou}
\end{align}
\par\noindent
Moreover, $4$ expressions of $\theta_{s}$, that is, Eqs. \eqref{onene12sou}, \eqref{onene22sou},
\eqref{onene32sou}, and \eqref{onene42sou} suggest
\begin{eqnarray*}
\theta_{s}
\!\!\!&=&\!\!\!- \dfrac{\sqrt{2}\lambda^{2}\alpha +  e^{i \sigma_{+}} \beta}{\lambda\alpha}=\dfrac{ \beta}{ \lambda(e^{-i \sigma_{+}} \alpha + \sqrt{2} \beta)}\\
\!\!\!&=&\!\!\!\dfrac{\alpha}{\lambda(\sqrt{2}\alpha-\beta e^{i \sigma_{-}})}=\dfrac{e^{-i\sigma_{-}}\alpha-\sqrt{2} \lambda^{2} \beta}{\lambda\beta}.
\end{eqnarray*}
The above equations give the combinations of $\lambda^{(j)}$ and $\Psi^{(j)}(0)={}^T\![\alpha^{(j)},\beta^{(j)}]\;(j=1,2,3,4)$ in Proposition \ref{prob-meas.} and Theorem \ref{stat}.
Here, the proof of Proposition \ref{prob-meas.} is complete.
Noting that the stationary measure is defined by $\mu(x)=|\Psi^{R}(x)|^{2}+|\Psi^{L}(x)|^{2}\;(x\in\mathbb{R})$, we obtain Theorem $1$.


\section{Proof of Theorem \ref{average}}
\label{proof-theo2}
At first, we introduce some notations.
Here we should note that $U_{x}$ can be divided into two parts as 
\[U_{x}=P_{x}+Q_{x},\]
where
\[ 
P_{x}=\begin{bmatrix} 
a_{x} & b_{x} \\
0 & 0 
\end{bmatrix}, \;\;
Q_{x}=\begin{bmatrix}
0 & 0 \\
c_{x} & d_{x}
\end{bmatrix}.\]
Let $\Xi_{t}(l,m)$ be the weight of all the passages of the QW which moves left $l$ times and moves right $m$ times till time $t\:\cite{segawa}$;
\[\Xi_{t}(l,m)=\sum_{l_{j},m_{j}}P_{x_{l1}}^{l_{1}}Q_{x_{m1}}^{m_{1}}P_{x_{l2}}^{l_{2}}Q_{x_{m_{2}}}^{m_{2}}\cdots P_{x_{lt}}^{l_{t}}Q_{x_{mt}}^{m_{t}}.\]\\
Note $l+m=t,\;-l+m=x,\;\;\sum_{i}l_{i}=l,\;\sum_{j}m_{j}=m$, and $\;\sum_{\gamma=l_{i},m_{j}}|x_{\gamma}|=x$. 
Now we present a useful fact to investigate localization. The time-averaged limit measure is written by the
square norm of the residue of the generating function $\tilde{\Xi}_{x}(z)\equiv\sum_{t\geq0}\Xi_{t}(x)z^{t}$ as follows;
\par\indent
\par\noindent
\begin{prop}\label{monsan}$\cite{endo}$
We have
\[\overline{\mu}_{\infty}(x)=\sum_{\theta_{s}}\|Res(\tilde{\Xi}_{x}(z);z=e^{i\theta_{s}})\varphi_{0}\|^{2},\]
where $\{e^{i\theta_{s}}\}$ is the set of the singular points of $\tilde{\Xi}_{x}(z)$.
\end{prop}
\par\indent
\par\noindent
We should note that we consider $z\in\mathbb{C}$ with $|z|=1$.
Here we give useful concrete expressions of $\tilde{\Xi}_{x}(z)$. 
The expressions play important roles in the proof. 
The derivation of Lemma \ref{kls} comes from a direct interpretation to our two-phase QW from Lemma $3.1$ 
in Ref. \cite{segawa}. Assume that the coin starts at the origin with the initial state $\varphi_{0}={}^T\![\alpha,\beta]$ with $\alpha,\beta\in\mathbb{C}$,
and $|\alpha|^{2}+|\beta|^{2}=1$.
\par\indent

\begin{lemma} \label{kls} 
\begin{enumerate}
\item If $x=0$, we have 
\[\tilde{\Xi}_{0}(z)=\dfrac{1}{1+\tilde{f}_{0}^{(+)}(z)\tilde{f}_{0}^{(-)}(z)}
\begin{bmatrix} 
1 & -\tilde{f}_{0}^{(+)}(z)\\
\tilde{f}_{0}^{(-)}(z) & 1\\
\end{bmatrix}.\]
\item If $|x|\geq 1$, we have
\[
\tilde{\Xi}_{x}(z)=\left\{\begin{array}{ll}
(\tilde{\lambda}^{(+)}(z))^{x-1}
\left[
    \begin{array}{c}
      \tilde{\lambda}^{(+)}(z)\tilde{f}_{0}^{(+)}(z)\\
      z \\
    \end{array}
  \right][0,-1]\tilde{\Xi}_{0}(z) & (x\geq 1), \\
  &\\
(\tilde{\lambda}^{(-)}(z))^{|x|-1}
\left[
    \begin{array}{c}
      z \\
      \tilde{\lambda}^{(-)}(z)\tilde{f}_{0}^{(-)}(z) \\
    \end{array}
  \right][1,0]\tilde{\Xi}_{0}(z) & (x\leq -1), \\
\end{array} \right.\]
\end{enumerate}
\noindent
where $\tilde{\lambda}^{(+)}(z)=\dfrac{z}{e^{-i\sigma_{+}}\tilde{f}_{0}^{(+)}(z)-\sqrt{2}},\;\tilde{\lambda}^{(-)}(z)=\dfrac{z}{\sqrt{2}-e^{i\sigma_{-}}\tilde{f}_{0}^{(-)}(z)}.$ 
Here $\tilde{f}_{0}^{(+)}(z)$ and $\tilde{f}_{0}^{(-)}(z)$ satisfy the following quadratic equations;
\[
\left\{
\begin{array}{l}
(\tilde{f}^{(+)}_{0}(z))^{2}-\sqrt{2}e^{i\sigma_{+}}(1+z^{2})\tilde{f}^{(+)}_{0}(z)+e^{2i\sigma_{+}}z^{2}=0,\\
\\
(\tilde{f}^{(-)}_{0}(z))^{2}-\sqrt{2}e^{-i\sigma_{-}}(1+z^{2})\tilde{f}^{(-)}_{0}(z)+e^{-2i\sigma_{-}}z^{2}=0.
\end{array}
\right.
\]
\end{lemma}
Therefore, we obtain
\par\indent
\par\noindent
\begin{lemma}
\label{monpal}
$\tilde{f}_{0}^{(+)}(z)$ and $\tilde{f}_{0}^{(-)}(z)$ are expressed in terms of $\theta$ by
\begin{align}\tilde{f}_{0}^{(\pm)}(z)=e^{i(\theta\pm\sigma_{\pm})}\times e^{i\tilde{\phi}(\theta)},\end{align}
where 
\begin{align}\left\{
\begin{array}{l}
\sin\tilde{\phi}(\theta)=\operatorname{sgn}(\sin\theta)\sqrt{2\sin\theta^{2}-1},\\
\cos\tilde{\phi}(\theta)=\sqrt{2}\cos\theta.
\end{array}
\right.\label{tildephi}\end{align} 
\end{lemma}
\par\indent
\par\noindent
Then, taking advantage of Lemma \ref{kls}, we get the set of the singular points of $\tilde{\Xi}_{x}(z)$ as follows;
\par\indent
\par\noindent
\begin{lemma}
\label{kai}
Let
\[e^{i\theta^{(\pm)}_{1}}=\pm\left(\dfrac{\cos\sigma}{\sqrt{3-2\sqrt{2}\sin\sigma}}+\dfrac{\sqrt{2}-\sin\sigma}{\sqrt{3-2\sqrt{2}\sin\sigma}}i\right),
\;\;\;e^{i\theta^{(\pm)}_{2}}=\pm\left(\dfrac{\cos\sigma}{\sqrt{3+2\sqrt{2}\sin\sigma}}-\dfrac{\sqrt{2}+\sin\sigma}{\sqrt{3+2\sqrt{2}\sin\sigma}}i\right),\]
with $|z|=1$.
Then, the set of all the singular points of $\;\tilde{\Xi}_{x}(z)$ with $|z|=1$, B, is given by
\begin{eqnarray*}
B=\left\{ \begin{array}{ll}
B_{1}=\{e^{i\theta^{(+)}_{1}},e^{i\theta^{(-)}_{1}}\},& \rm{if} -1\leq\sin\sigma\leq1/\sqrt{2},\\
B_{2}=\{e^{i\theta^{(+)}_{1}},e^{i\theta^{(-)}_{1}},e^{i\theta^{(+)}_{2}},e^{i\theta^{(-)}_{2}}\},& \rm{if} -1/\sqrt{2}\leq\sin\sigma\leq1/\sqrt{2},\\
B_{3}=\{e^{i\theta^{(+)}_{2}},e^{i\theta^{(-)}_{2}}\},& \rm{if}\;-1/\sqrt{2}\leq\sin\sigma\leq1.\\
\end{array} \right.
\end{eqnarray*} 
\end{lemma}
We give the proof of Lemma \ref{kai} in Appendix B. \\
Then, we derive the residues of $\;\tilde{\Xi}_{x}(z)$ at the singular points. 
Noting that all the singular points of come from the denominator of $\tilde{\Xi}_{0}(z)$, and we put $\tilde{\Lambda}_{0}(z)\equiv1+\tilde{f}^{(+)}_{0}(z)\tilde{f}^{(-)}_{0}(z)$. 
As we can see in Appendix B, every singular point $z_{s}\in B$ is derived from the solution of 
\[\tilde{\Lambda}_{0}(z)=0.\]
Then, Lemma \ref{pari} gives explicit expressions of the square of the absolute value of the residues of $1/\tilde{\Lambda}_{0}(z)$ as follows. The proof is given in Appendix C.

\par\indent
\begin{lemma}
\label{pari}
\begin{enumerate}
\item For $e^{i\theta^{(\pm)}_{1}}$, we have
\[\left|Res\left(\dfrac{1}{\tilde{\Lambda}_{0}(z)}; z=e^{i\theta_{1}^{(\pm)}}\right)\right|^{2}=\dfrac{1}{4}\left|\dfrac{\sqrt{2}\sin\sigma-1}{2\sqrt{2}\sin\sigma-3}\right|^{2}.\]
\item For $e^{i\theta^{(\pm)}_{2}}$, we have
\[\left|Res\left(\dfrac{1}{\tilde{\Lambda}_{0}(z)}; z=e^{i\theta_{2}^{(\pm)}}\right)\right|^{2}=\dfrac{1}{4}\left|\dfrac{\sqrt{2}\sin\sigma+1}{2\sqrt{2}\sin\sigma+3}\right|^{2}.\]
\end{enumerate}
\end{lemma}

\par\indent
\par\noindent
Owing to Lemma $2$-$5$, we prove first the case of $x=0$ in Theorem $2$. According to Lemma \ref{kls}, we have
\[\tilde{\Xi}_{0}(z)\varphi_{0}=\dfrac{1}{\tilde{\Lambda}_{0}(z)}\begin{bmatrix}\alpha-\beta\tilde{f}_{0}^{(+)}(z)\\ \alpha\tilde{f}_{0}^{(-)}(z)+\beta\end{bmatrix}.\]
Then, we obtain the square norm of the residues;
\begin{align}\|Res(\tilde{\Xi}_{0}(z)\varphi_{0}:z=e^{i\theta})\|^{2}=\left|Res\left(\dfrac{\alpha-\beta\tilde{f}_{0}^{(+)}(z)}{\tilde{\Lambda}_{0}(z)}:z=e^{i\theta}\right)\right|^{2}+\left|Res\left(\dfrac{\alpha\tilde{f}_{0}^{(-)}(z)+\beta}{\tilde{\Lambda}_{0}(z)}:z\right)\right|^{2}.\label{squarenorm-residue}\end{align}
Noting that $Res(1/\tilde{\Lambda}_{0}(z):z=e^{i\theta})=\lim_{z\to e^{i\theta}}(z-e^{i\theta})/\tilde{\Lambda}_{0}(z) $ holds for any $\theta$, we see by expanding $\tilde{\Lambda}_{0}(z)$ around the solution $e^{i\theta}$,
\begin{align}\left|Res\left(\dfrac{1}{\tilde{\Lambda}_{0}(z)}:z=e^{i\theta}\right)\right|^{2}=\dfrac{1}{|\tilde{\Lambda}^{'}_{0}(e^{i\theta})|^{2}}=\dfrac{1}{4\left|1+\dfrac{\partial\tilde{\phi}(\theta)}{\partial\theta}\right|^{2}},\label{expanding-res}\end{align}
where
\[\tilde{\Lambda}^{'}_{0}(z)=\dfrac{\partial\tilde{\Lambda}_{0}(z)}{\partial z}.\]
Equation \eqref{expanding-res} gives
\[
\left\{
\begin{array}{l}
\left|Res\left(\dfrac{\alpha-\beta\tilde{f}_{0}^{(+)}(z)}{\tilde{\Lambda}_{0}(z)}:z=e^{i\theta}\right)\right|^{2}=\dfrac{|\alpha-\beta\tilde{f}_{0}^{(+)}(e^{i\theta})|^{2}}{4\left|1+\dfrac{\partial\tilde{\phi}(\theta)}{\partial\theta}\right|^{2}},\\
\left|Res\left(\dfrac{\alpha\tilde{f}_{0}^{(-)}(z)+\beta}{\tilde{\Lambda}_{0}(z)}:z=e^{i\theta}\right)\right|^{2}=\dfrac{|\alpha\tilde{f}_{0}^{(-)}(e^{i\theta})+\beta|^{2}}{4\left|1+\dfrac{\partial\tilde{\phi}(\theta)}{\partial\theta}\right|^{2}}.
\end{array}\right.\]
\noindent
Hence, we see
\[\|Res(\tilde{\Xi}_{0}(z)\varphi_{0}:z=e^{i\theta})\|^{2}=\dfrac{1}{2\left|1+\dfrac{\partial\tilde{\phi}(\theta)}{\partial\theta}\right|^{2}}\{1-\Re(\overline{\alpha}\beta\tilde{f}_{0}^{(+)}(e^{i\theta}))+\Re(\alpha\overline{\beta}\tilde{f}_{0}^{(-)}(e^{i\theta}))\},\]
which implies
\begin{align}\overline{\mu}_{\infty}(0)=\sum_{\theta}\dfrac{1}{2\left|1+\dfrac{\partial\tilde{\phi}(\theta)}{\partial\theta}\right|^{2}}\{1-\Re(\overline{\alpha}\beta\tilde{f}_{0}^{(+)}(e^{i\theta}))+\Re(\alpha\overline{\beta}\tilde{f}_{0}^{(-)}(e^{i\theta}))\}\label{timeaverage-dainyuu}\end{align}
where the range of the summation is 
\[\{\theta\in[0,2\pi); e^{i\theta}\in B\}.\]
Note that $\Re(z)$ expresses the real part of $z$. \\
\par\indent
\par\noindent
{\bf Remark :} 
From Lemma \ref{monpal}, we get
\begin{enumerate}
\item $\Re{[\overline{\alpha}\beta\tilde{f}^{(+)}_{0}(e^{i\theta_{1}^{(\pm)}})]}=-ab\sin(\tilde{\sigma}-\tilde{\phi}_{12})$,
\item $\Re{[\alpha\overline{\beta}\tilde{f}^{(-)}_{0}(e^{i\theta_{1}^{(\pm)}})]}=ab\sin(\tilde{\sigma}-\tilde{\phi}_{12})$,
\item $\Re{[\overline{\alpha}\beta\tilde{f}^{(+)}_{0}(e^{i\theta_{2}^{(\pm)}})]}=ab\sin(\tilde{\sigma}-\tilde{\phi}_{12})$,
\item $\Re{[\alpha\overline{\beta}\tilde{f}^{(-)}_{0}(e^{i\theta_{2}^{(\pm)}})]}=-ab\sin(\tilde{\sigma}-\tilde{\phi}_{12})$.
\end{enumerate}
\par\indent
\par\noindent
Noting Lemma \ref{pari} and substituting the computed items of 
$\Re{[\overline{\alpha}\beta\tilde{f}^{(\pm)}_{0}(e^{i\theta_{j}^{(\pm)}})]}\;(j=1,2)$ into Eq. \eqref{timeaverage-dainyuu}, we obtain the desired conclusion for the case of $x=0$ in Theorem \ref{average}.\\
\indent
Next, we give the proof for the case of $x\geq1$ in Theorem \ref{average}.
From Lemma \ref{kls}, we have
\[\tilde{\Xi}_{x}(z)\varphi_{0}=-\dfrac{(\tilde{\lambda}^{(+)}(z))^{x-1}}{\tilde{\Lambda}_{0}(z)}\begin{bmatrix}\tilde{\lambda}^{(+)}(z)\tilde{f}_{0}^{(+)}(z)(\alpha\tilde{f}_{0}^{(-)}(z)+\beta)\\ z(\alpha\tilde{f}_{0}^{(-)}(z)+\beta)\end{bmatrix}\; (x\geq 1).\]
Then, we obtain the square norm of the residues by\\
\par\noindent
$\|Res(\tilde{\Xi}_{x}(z)\varphi_{0}:z=e^{i\theta})\|^{2}=\left|Res\left(-\dfrac{(\tilde{\lambda}^{(+)}(z))^{x}\tilde{f}_{0}^{(+)}(e^{i\theta})(\alpha\tilde{f}_{0}^{(-)}(z)+\beta)}{\tilde{\Lambda}_{0}(z)}:z=e^{i\theta}\right)\right|^{2}$

\begin{align}+\left|Res\left(-\dfrac{(\tilde{\lambda}^{(+)}(z))^{x-1}z(\alpha\tilde{f}_{0}^{(-)}(z)+\beta))}{\tilde{\Lambda}_{0}(z)}:z=e^{i\theta}\right)\right|^{2}.\label{norm}\end{align}
Noting
\[\;Res\left(-\dfrac{(\tilde{\lambda}^{(+)}(z))^{x}\tilde{f}_{0}^{(+)}(z)(\alpha\tilde{f}_{0}^{(-)}(z)+\beta)}{\tilde{\Lambda}_{0}(z)}:z=e^{i\theta}\right)=-\dfrac{(\tilde{\lambda}^{(+)}(e^{i\theta}))^{x}\tilde{f}_{0}^{(+)}(e^{i\theta})(\alpha\tilde{f}_{0}^{(-)}(e^{i\theta})+\beta)}{\tilde{\Lambda}^{'}_{0}(e^{i\theta})},\]
we have
\begin{align}\left|Res\left(-\dfrac{(\tilde{\lambda}^{(+)}(z))^{x}\tilde{f}_{0}^{(+)}(z)(\alpha\tilde{f}_{0}^{(-)}(z)+\beta)}{\tilde{\Lambda}_{0}(z)}:z=e^{i\theta}\right)\right|^{2}=\dfrac{|\tilde{\lambda}^{(+)}(e^{i\theta})|^{2x}|(\alpha\tilde{f}_{0}^{(-)}(e^{i\theta})+\beta)|^{2}}{4\left|1+\dfrac{\partial\tilde{\phi}(\theta)}{\partial\theta}\right|^{2}}.\label{ookisa1}\end{align}
In the similar way, we get
\begin{align}\left|Res\left(-\dfrac{(\tilde{\lambda}^{(+)}(z))^{x-1}z(\alpha\tilde{f}_{0}^{(-)}(z)+\beta)}{\tilde{\Lambda}_{0}(z)}:z=e^{i\theta}\right)\right|^{2}=\dfrac{|\tilde{\lambda}^{(+)}(e^{i\theta})|^{2(x-1)}|(\alpha\tilde{f}_{0}^{(-)}(e^{i\theta})+\beta)|^{2}}{4\left|1+\dfrac{\partial\tilde{\phi}(\theta)}{\partial\theta}\right|^{2}}.\label{ookisa2}\end{align}
Hence,  Eqs. \eqref {norm}, \eqref{ookisa1} and \eqref{ookisa2} give for $\theta\in\mathbb{R}$, 
\[\|Res(\tilde{\Xi}_{x}(z)\varphi_{0}:z=e^{i\theta})\|^{2}=\dfrac{|\tilde{\lambda}^{(+)}(e^{i\theta})|^{2(x-1)}}{4\left|1+\dfrac{\partial\tilde{\phi}(\theta)}{\partial\theta}\right|^{2}}(1+|\tilde{\lambda}^{(+)}(e^{i\theta})|^{2})\{1+2\Re{(\alpha\overline{\beta}\tilde{f}_{0}^{(-)}(e^{i\theta}))}\}.\]
So we obtain
\begin{align}\overline{\mu}_{\infty}(x)=\sum_{\theta_{s}}\dfrac{|\tilde{\lambda}^{(+)}(e^{i\theta_{s}})|^{2(x-1)}}{4\left|1+\dfrac{\partial\tilde{\phi}(\theta)}{\partial\theta}\right|^{2}}(1+|\tilde{\lambda}^{(+)}(e^{i\theta_{s}})|^{2})\{1+2\Re{(\alpha\overline{\beta}\tilde{f}_{0}^{(-)}(e^{i\theta_{s}}))}\},\label{jikanheikin}\end{align}
where $\{e^{i\theta_{s}}\}$ is the set of the singular points for $\tilde{\Xi}_{x}(z).$
Next, we compute $|\tilde{\lambda}^{(+)}(e^{i\theta_{s}})|^{2}$.
The definition of $\tilde{\lambda}^{(+)}(z)$ in Lemma \ref{kls} gives 
\[\tilde{\lambda}^{(+)}(e^{i\theta})=\dfrac{1}{e^{i\tilde{\phi}(\theta)}-\sqrt{2}e^{-i\theta}}.\]
Hence we see
\[|\tilde{\lambda}^{(+)}(e^{i\theta})|^{2}=\dfrac{1}{3-2\sqrt{2}\cos(\theta+\tilde{\phi}(\theta))},\]
Equation (B.2) in Appendix B implies
\begin{eqnarray}\left\{ \begin{array}{ll}
|\tilde{\lambda}^{(+)}(e^{i\theta^{(\pm)}_{1}})|^{2}=\dfrac{1}{3-2\sqrt{2}\sin\sigma}&-1\leq\sin\sigma\leq1/\sqrt{2},\\
|\tilde{\lambda}^{(+)}(e^{i\theta^{(\pm)}_{2}})|^{2}=\dfrac{1}{3+2\sqrt{2}\sin\sigma}&-1//\sqrt{2}\leq\sin\sigma\leq1.
\end{array} \right.\label{lambdaookisa}\end{eqnarray} 
Noting Proposition \ref{monsan} and Lemma \ref{pari}, and substituting the computed items of 
$\Re{[\overline{\alpha}\beta\tilde{f}^{(\pm)}_{0}(e^{i\theta_{j}^{(\pm)}})]}\;(j=1,2)$ in ``${\it \bf Remark}$'' and Eq. \eqref{lambdaookisa} into Eq. \eqref{jikanheikin}, we obtain the case of $x\geq1$ in Theorem $2$.
In a similar way, we also get the case of $x\leq-1$ in Theorem $2$, which completes the proof of Theorem $2$.
\section{Summary}
In this paper, we presented two theorems concerning
localization, that is, the stationary and
time-averaged limit measures for the two-phase QW with one defect. We showed that
localization can be observed regardless of
the parameters of the model and initial state, though the
time-averaged limit measure heavily
depends on the parameters of the model and initial state.
When the parameters have specific periodicities, and assuming the
corresponding relationships
between $\lambda^{(j)} (j = 1, 2, 3, 4)$ in Theorem \ref{stat} and the range of
$\sin\sigma$,
we found that  there is an interesting relationship between the
stationary and time-averaged
limit measures as we gave two examples in Subsection \ref{examples}.
Moreover, we showed that the probability distribution does not have
an origin symmetry (see Figs. \ref{fig3}, \ref{fig4}),
however, the time-averaged limit measure has the symmetry. In the
forthcoming paper, we will report that the
asymmetry of the probability distribution can be expressed by the weak
limit theorem for $X_t/t$.

\setcounter{footnote}{0}
\renewcommand{\thefootnote}{\alph{footnote}}
\nonumsection{Acknowledgments}
SE acknowledges financial support of Postdoctoral Fellowship for Research Abroad from Japan Society for the Promotion of Science.
NK acknowledges financial support of the Grant-in-Aid for Scientific
Research (C) of Japan Society for the Promotion of Science (No.21540116). 
ES acknowledges the partial support by the Grant-in-Aid for young Scientists (B) of Japan Society for the promotion of Science (No.25800088).\\
\par\noindent
\nonumsection{References}
\begin{appendix}\\ 
In Appendix A, we explain how $\tilde{f}_{x}^{(\pm)}(z)$ and $\tilde{\lambda}^{(\pm)}$ are determined. 
At first, we should remark that by the quadratic equations of $\tilde{f}^{(\pm)}_{0}(z)$ and the definitions of $\tilde{\lambda}^{(\pm)}(z)$ in Lemma \ref{kls}, we see 
\begin{eqnarray}
\left\{ \begin{array}{ll}
\tilde{f}_{x}^{(+)}(z)=\tilde{f}_{0}^{(+)}(z),\;\tilde{\lambda}^{(+)}_{x}(z)=\tilde{\lambda}^{(+)}(z)& (x\geq1), \\
\tilde{f}_{x}^{(-)}(z)=\tilde{f}_{0}^{(-)}(z),\;\tilde{\lambda}^{(-)}_{x}(z)=\tilde{\lambda}^{(-)}(z)&(x\leq-1),
\end{array} \right.
\end{eqnarray}
where $\tilde{\lambda}^{(+)}=zd_{+}/(1-c_{+}\tilde{f}_{0}^{(+)}(z))\;\;$ and $\;\;\tilde{\lambda}^{(-)}(z)=za_{-}/(1-b_{-}\tilde{f}_{0}^{(-)}(z))$.\\
We should also note that $\tilde{f}^{(+)}_{0}(z)$ and $\tilde{f}^{(-)}_{0}(z)$ are one of the solutions of
\begin{eqnarray}
\left\{ \begin{array}{ll}
(\tilde{f}^{(+)}_{0}(z))^{2}+\sqrt{2}e^{i\sigma_{+}}w(w-w^{-1})\tilde{f}^{(+)}_{0}(z)-e^{2i\sigma_{+}}w^{2}=0,
&\\
(\tilde{f}^{(-)}_{0}(z))^{2}+\sqrt{2}e^{-i\sigma_{-}}w(w-w^{-1})\tilde{f}^{(-)}_{0}(z)-e^{-2i\sigma_{-}}w^{2}=0,
\end{array} \right.
\end{eqnarray}
respectively, where $\;w=iz=ie^{i\theta}$.\\
Hence, we have 
\[
\left\{
\begin{array}{l}
\tilde{\lambda}^{(\pm)}(w)=\pm\dfrac{i}{\sqrt{2}}\{(w+w^{-1})-\sqrt{(w+w^{-1})^{2}-2}\},\\
\tilde{f}^{(\pm)}_{0}(w)=-\dfrac{w e^{i\sigma_{+}}}{\sqrt{2}}\{(w-w^{-1})+\sqrt{(w-w^{-1})^{2}+2}\}.
\end{array}
\right.\]

Putting $w=i(1-\epsilon)e^{i\theta}\;(\epsilon\in\mathbb{R},\;|\epsilon|\ll1)$, we show how $\lim_{\epsilon\to0}\sqrt{(w+w^{-1})^{2}-2}$ can be written in terms of $\theta$ according to the range of $\cos\theta$ or $\sin\theta$.
Noting $|\epsilon|\ll1$, we can approximates $\tilde{\lambda}^{(\pm)}(w)$ as follows;
\begin{eqnarray}
\tilde{\lambda}^{(\pm)}(w)&\!\!\!=\!\!\!&\pm\dfrac{i}{\sqrt{2}}\left\{(1-\epsilon)ie^{i\theta}-(1-\epsilon)^{-1}ie^{-i\theta}-\sqrt{\{(1-\epsilon)ie^{i\theta}-(1-\epsilon)^{-1}ie^{-i\theta}\}^{2}-2}\right\}\nonumber\\
&\sim\!\!\!&\dfrac{i}{\sqrt{2}}\left\{-2\sin\theta-2i\epsilon\cos\theta-\delta\sqrt{4\sin^{2}\theta-2}\right\}\nonumber\\
&=\!\!\!&\mp\dfrac{i}{\sqrt{2}}\left\{2\sin\theta+2i\epsilon\cos\theta+\delta\sqrt{4\sin^{2}\theta-2}\right\}.\label{kuma}\end{eqnarray}
Here we put $\delta\in\mathbb{R}$ with $\delta^{2}=1$. 
Now, we chose the square root so that $|\tilde{\lambda}^{(\pm)}(w)|<1$ for $|z|<1$.
Equation (\ref{kuma}) leads to the following two cases;
\begin{enumerate}
\item Case of $|\sin\theta|\geq 1/\sqrt{2}$:\\
Equation (\ref{kuma}) suggests  
\[\dfrac{1}{2}\left\{2\sin\theta+2\delta\sqrt{\sin^{2}\theta-1/2}\right\}^{2}<1,\]
which leads to
\[2\sin^{2}\theta+2\sin\theta\delta\sqrt{\sin^{2}\theta-1/2}<1.\]
Therefore, we obtain $\delta=-\operatorname{sgn}(\sin\theta)$.
\item Case of $|\sin\theta|<1/\sqrt{2}$:\\
Equation (\ref{kuma}) also suggests 
\[\dfrac{1}{2}\left[\left\{2\sin\theta+2\delta\sqrt{\sin^{2}\theta-1/2}\right\}^{2}+4\epsilon^{2}\cos^{2}\theta\right]<1,\]
which implies
\[4\epsilon^{2}\cos^{2}\theta+8\epsilon\cos\theta\delta\sqrt{1/2-\sin^{2}\theta}<0.\]
Therefore, we obtain $\delta=-\operatorname{sgn}(\cos\theta)$.
\end{enumerate}
As a result, the square root is expressed as follows;
\begin{align}\lim_{\epsilon\to0}\sqrt{(w+w^{-1})^{2}-2}=\left\{
\begin{array}{ll}
-2\operatorname{sgn}(\sin\theta)\sqrt{\sin^{2}\theta-\dfrac{1}{2}}&(\;|\sin\theta|\geq1/\sqrt{2}\;),\\
-2i\operatorname{sgn}(\cos\theta)\sqrt{\dfrac{1}{2}-\sin^{2}\theta}&(\;|\sin\theta|\leq1/\sqrt{2}\;).
\end{array}
\right.\end{align}

Next, we consider $\tilde{\lambda}^{(\pm)}(z)$ and $\tilde{f}_{0}^{(\pm)}(z)$ in detail.
If we focus on localization of the two-phase QW with one defect, we choose the square root so that $1/\tilde{\Lambda}_{0}(z)=1+\tilde{f}^{(+)}_{0}(z)\tilde{f}^{(-)}_{0}(z)$ has the singular points, in other words, $|\tilde{f}^{(\pm)}_{0}(z)|=1$.
The above discussion implies in this case,
\begin{align}\left\{
\begin{array}{l}
\tilde{\lambda}^{(\pm)}(z)=\mp i\{\sqrt{2}\sin\theta-\operatorname{sgn}(\sin\theta)\sqrt{1-2\cos^{2}\theta}\}\\
\tilde{f}^{(\pm)}_{0}(z)=e^{i(\theta\pm\sigma_{\pm}+\tilde{\phi}(\theta))}
\end{array}
\right.\;(|\sin\theta|\geq 1/\sqrt{2}),\end{align}
where $z=e^{i\theta}$, and $\tilde{\phi}(\theta)$ is defined by Eq. \eqref{tildephi}.
\end{appendix}
\flushleft
\begin{appendix}
\noindent
In Appendix B, we provide with the proof of Lemma \ref{kai}. The aim of Appendix B is to derive all the singular points of $\tilde{\Xi}_{x}(z)$ for the two-phase QW with one defect.
We should recall that the singular points of $\tilde{\Xi}_{x}(z)$ come from $1/\tilde{\Lambda}_{0}(z)$ part of  $\tilde{\Xi}_{x}(z)$. Here, we should note the expressions $\tilde{\Lambda}_{0}(e^{i\theta})=1+\tilde{f}_{0}^{(+)}(e^{i\theta})\tilde{f}_{0}^{(-)}(e^{i\theta})$ and $\tilde{f}_{0}^{(\pm)}(e^{i\theta})=e^{i(\theta\pm\sigma_{\pm})}\times e^{i\tilde{\phi}(\theta)}$, 
where $\tilde{\phi}(\theta)$ is defined by Eq. \eqref{tildephi}. 
Hence we have
\[\tilde{\Lambda}_{0}(e^{i\theta})=1+e^{2i(\theta+\sigma+\tilde{\phi}(\theta))},\]
where
$\;\sigma=(\sigma_{+}-\sigma_{-})/2.\;$ Thus, we need to derive all $\theta$ satisfying
\begin{align}\tilde{\Lambda}_{0}(e^{i\theta})=1+e^{2i(\theta+\sigma+\tilde{\phi}(\theta))}=0.\label{butasan}\end{align}
Equation (\ref{butasan}) yields $e^{i(\theta+\tilde{\phi}(\theta))}=ie^{-i\sigma}$ or $-ie^{-i\sigma}$.\\
\par\indent
\par\noindent
$(1)$ Case of $\;e^{i(\theta+\tilde{\phi}(\theta))}=ie^{-i\sigma}$:\\
Noting
\begin{align}
\left\{
\begin{array}{l}
\cos(\theta+\tilde{\phi}(\theta))=\sin\sigma,\\
\sin(\theta+\tilde{\phi}(\theta))=\cos\sigma,\\
\end{array}
\right.\label{usi}\end{align}
we have the solutions of Eq. \eqref{butasan} as follows:
\begin{align}\sin^{2}\theta_{1}=\dfrac{(\sqrt{2}-\sin\sigma)^{2}}{3-2\sqrt{2}\sin\sigma},\;\;\;\cos^{2}\theta_{1}=\dfrac{\cos^{2}\sigma}{3-2\sqrt{2}\sin\sigma}.\label{tori}\end{align}
By simple observations for Eq. \eqref{tori}, we see that when $-1\leq\sin\sigma\leq1/\sqrt{2}$, the following relations hold.
\begin{enumerate}
\item $\operatorname{sgn}(\cos\sigma)=1$ case;\\

\[
\left\{
\begin{array}{l}
\operatorname{sgn}(\sin\theta_{1})=\pm1,\\
\operatorname{sgn}(\cos\theta_{1})=\pm1.\\
\end{array}
\right.
\] 
\item  $\operatorname{sgn}(\cos\sigma)=-1$ case;\\
\[
\left\{
\begin{array}{l}
\operatorname{sgn}(\sin\theta_{1})=\pm1,\\
\operatorname{sgn}(\cos\theta_{1})=\mp1.\\
\end{array}
\right.
\]
\end{enumerate}
$(2)$ Case of $e^{i(\theta+\tilde{\phi})}=-ie^{-i\sigma}$:\\
Noting
\begin{align}
\left\{
\begin{array}{l}
\cos(\theta+\tilde{\phi}(\theta))=-\sin\sigma,\\
\sin(\theta+\tilde{\phi}(\theta))=-\cos\sigma,\\
\end{array}
\right.\label{wata}\end{align}
we see
\[\sin^{2}\theta_{2}=\dfrac{(\sqrt{2}+\sin\sigma)^{2}}{3+2\sqrt{2}\sin\sigma},\;\;\;\cos^{2}\theta_{2}=\dfrac{\cos^{2}\sigma}{3+2\sqrt{2}\sin\sigma}.\]
From Eq. \eqref{wata}, we see that when $-1/\sqrt{2}\leq\sin\sigma\leq1$, we get
\begin{enumerate}
\item $\operatorname{sgn}(\cos\sigma)=1$ case;\\

\[
\left\{
\begin{array}{l}
\operatorname{sgn}(\sin\theta_{1})=\pm1,\\
\operatorname{sgn}(\cos\theta_{1})=\mp1.\\
\end{array}
\right.
\] 
\item $\operatorname{sgn}(\cos\sigma)=-1$ case;\\
\[
\left\{
\begin{array}{l}
\operatorname{sgn}(\sin\theta_{1})=\pm1,\\
\operatorname{sgn}(\cos\theta_{1})=\pm1.\\
\end{array}
\right.
\]
\end{enumerate}
We summerize the above discussion in the following table;\\
\par\noindent

\begin{table}[!h]
  \begin{tabular}{|c|c|c|c|} \hline
    range of $\sin\sigma$ &  $\cos\theta$& $\sin\theta$ & complex number form\\ \hline 
  $-1\leq\sin\sigma\leq 1/\sqrt{2}$ &$ \pm\cos\sigma/\sqrt{3-2\sqrt{2}\sin\sigma}$ & $\pm(\sqrt{2}-\sin\sigma)/\sqrt{3-2\sqrt{2}\sin\sigma}$ & $e^{i\theta_{1}^{(\pm)}}$ \\ \hline   
    $-1/\sqrt{2}\leq\sin\sigma\leq1$ & $\mp\cos\sigma/\sqrt{3+2\sqrt{2}\sin\sigma}$ & $\pm(\sqrt{2}+\sin\sigma)/\sqrt{3+2\sqrt{2}\sin\sigma}$ & $e^{i\theta_{2}^{(\pm)}}$\\ \hline
    $-1/\sqrt{2}\leq\sin\sigma$ &$ \pm\cos\sigma/\sqrt{3-2\sqrt{2}\sin\sigma}$, 
    &$\pm(\sqrt{2}-\sin\sigma)/\sqrt{3-2\sqrt{2}\sin\sigma}$,   & $e^{i\theta_{1}^{(\pm)}}$, \\ 
    $\leq1/\sqrt{2}$& $\mp\cos\sigma/\sqrt{3+2\sqrt{2}\sin\sigma}$ &$\pm(\sqrt{2}+\sin\sigma)/\sqrt{3+2\sqrt{2}\sin\sigma}$&  $e^{i\theta_{2}^{(\pm)}}$
    \\\hline
  \end{tabular}
\end{table}

Here we put
\begin{itemize}
 \item $e^{i\theta_{1}^{(+)}}=\cos\sigma/\sqrt{3-2\sqrt{2}\sin\sigma}+i(\sqrt{2}-\sin\sigma)/\sqrt{3-2\sqrt{2}\sin\sigma}$,\\
 \item $e^{i\theta_{1}^{(-)}}=-\cos\sigma/\sqrt{3-2\sqrt{2}\sin\sigma}-i(\sqrt{2}-\sin\sigma)/\sqrt{3-2\sqrt{2}\sin\sigma}$,\\
 \item $e^{i\theta_{2}^{(+)}}=\cos\sigma/\sqrt{3+2\sqrt{2}\sin\sigma}-i(\sqrt{2}+\sin\sigma)/\sqrt{3+2\sqrt{2}\sin\sigma}$,\\
 \item $e^{i\theta_{2}^{(-)}}=-\cos\sigma/\sqrt{3+2\sqrt{2}\sin\sigma}+i(\sqrt{2}+\sin\sigma)/\sqrt{3+2\sqrt{2}\sin\sigma}$.
\end{itemize}
\end{appendix}
\flushleft
\begin{appendix}
\noindent
In Appendix C, we derive the norm of the residue of $\tilde{\Xi}_{0}(z)$ for the two-phase QW with one defect.
First of all, noting $Res\left(1/\tilde{\Lambda}_{0}(z):z=e^{i\theta}\right)=\lim_{z\to e^{i\theta}}(z-e^{i\theta})/\tilde{\Lambda}_{0}(z)$ and expanding $\tilde{\Lambda}_{0}(z)$ around $z=e^{i\theta}$, we have
\begin{align}\left|Res\left(\dfrac{1}{\tilde{\Lambda}_{0}(z)}:z=e^{i\theta}\right)\right|^{2}=\dfrac{1}{|\tilde{\Lambda}^{'}_{0}(e^{i\theta})|^{2}}.\label{nihon}\end{align}
Then, taking into account 
\[\tilde{\Lambda}^{'}_{0}(z)=\dfrac{\partial \tilde{\Lambda}_{0}(z)}{\partial z}=\dfrac{\partial \theta}{\partial z}\dfrac{\partial \tilde{\Lambda}_{0}(z)}{\partial \theta},
\]
and 
\[
\dfrac{\partial \theta}{\partial z}=-ie^{-i\theta},\;
\dfrac{\partial \tilde{\Lambda}_{0}(e^{i\theta})}{\partial \theta}=2i\left(1+\dfrac{\partial \tilde{\phi}}{\partial \theta}\right)e^{2i(\theta+\sigma+\tilde{\phi}(\theta))},
\]
we obtain
\begin{align}|\tilde{\Lambda}^{'}_{0}(e^{i\theta})|^{2}=4\left|1+\dfrac{\partial \tilde{\phi}(\theta)}{\partial \theta}\right|^{2}=4\left(1+\dfrac{\partial \tilde{\phi}(\theta)}{\partial \theta}\right)\left(1+\overline{\dfrac{\partial \tilde{\phi}(\theta)}{\partial \theta}}\right),\label{tokyo}\end{align}
where $\tilde{\phi}(\theta)$ is defined by Eq. \eqref{tildephi}.
Then, we derive the residues depending on the range of parameter $\sigma $.
Recall that $e^{i\theta_{1}^{(\pm)}}$ and $e^{i\theta_{2}^{(\pm)}}$ are the singular points of $\tilde{\Xi}_{0}(z)$ defined by Lemma \ref{kai}.
\begin{enumerate}
\item Case of $-1\leq\sin\sigma<1/\sqrt{2}$;\\
Let $B_{1}=\{e^{i\theta_{1}^{(+)}},e^{i\theta_{1}^{(-)}}\}$ be the set of the singular points of $\tilde{\Xi}_{0}(z)$ in this case.
Noting 
\[\dfrac{\partial \tilde{\phi}(\theta)}{\partial \theta}=\dfrac{\sqrt{2}|\sin\theta|}{\sqrt{1-2\cos^{2}\theta}},\]
we see 
\[\left.\dfrac{\partial \tilde{\phi}(\theta)}{\partial \theta}\right|_{\theta=\theta_{1}^{(\pm)}}=\dfrac{\sqrt{2}(\sqrt{2}-\sin\sigma)}{|1-\sqrt{2}\sin\sigma|}.\]
Therefore we have
\begin{align}4\left|1+\dfrac{\partial\tilde{\phi}(\theta)}{\partial\theta}\right|_{\theta=\theta_{1}^{(\pm)}}^{2}=4\left|\dfrac{3-2\sqrt{2}\sin\sigma}{1-\sqrt{2}\sin\sigma}\right|^{2}.\label{tohoku}\end{align}
Hence, Eqs. \eqref{nihon}, \eqref{tokyo} and Eq. \eqref{tohoku} give
\[\left|Res\left(\dfrac{1}{\tilde{\Lambda}_{0}(z)}:z=e^{i\theta_{1}^{(\pm)}}\right)\right|^{2}=\left\{\dfrac{1-\sqrt{2}\sin\sigma}{2(3-2\sqrt{2}\sin\sigma)}\right\}^{2}.\]

\item Case of $-1/\sqrt{2}<\sin\sigma\leq1$;\\
Let $B_{2}=\{e^{i\theta^{(+)}_{2}}, e^{i\theta^{(-)}_{2}}\}$ be the set of the singular points of $\tilde{\Xi}_{0}(z)$ in this case. \\
Noting
\[\dfrac{\partial \tilde{\phi}(\theta)}{\partial \theta}=\dfrac{\sqrt{2}|\sin\theta|}{\sqrt{2\sin\theta^{2}-1}},\]
we see 
\[\left.\dfrac{\partial \tilde{\phi}(\theta)}{\partial \theta}\right|_{\theta=\theta_{2}^{(\pm)}}=\dfrac{\sqrt{2}(\sqrt{2}+\sin\sigma)}{|1+\sqrt{2}\sin\sigma|}.\]
Therefore we have
\begin{align}4\left|1+\dfrac{\partial\tilde{\phi}(\theta)}{\partial\theta}\right|_{\theta=\theta_{2}^{(\pm)}}^{2}=4\left|\dfrac{3+2\sqrt{2}\sin\sigma}{1+\sqrt{2}\sin\sigma}\right|^{2}.\label{kansai}\end{align}
Now Eqs. \eqref{nihon}, \eqref{tokyo} and \eqref{kansai} imply
\[\left|Res\left(\dfrac{1}{\tilde{\Lambda}_{0}(z)}:z=e^{i\theta_{2}^{(\pm)}}\right)\right|^{2}=\left\{\dfrac{1+\sqrt{2}\sin\sigma}{2(3+2\sqrt{2}\sin\sigma)}\right\}^{2}.\]

\par\indent
\par\noindent
\end{enumerate}
By the above discussion, we obtain $\left|Res\left(1/\tilde{\Lambda}_{0}(z):z=e^{i\theta_{1}^{(\pm)}}\right)\right|^{2}$ and $\left|Res\left(1/\tilde{\Lambda}_{0}(z):z=e^{i\theta_{2}^{(\pm)}}\right)\right|^{2}$, and therefore we have completed the proof of Lemma \ref{pari}. As for the case of $-1/\sqrt{2}\leq\sin\sigma\leq1/\sqrt{2}$, all we need to consider is the norm of the residues for the singular points both $e^{i\theta_{1}^{(\pm)}}$ and $e^{i\theta_{2}^{(\pm)}}$.
\end{appendix}

\end{document}